\documentclass[preprint,12pt]{elsarticle}

\usepackage{amssymb}
\usepackage{amsmath}
\usepackage{lineno}
\usepackage{xcolor}
\usepackage{graphicx}
\usepackage{booktabs} 

\newproof{pf}{Proof}

\usepackage{amsmath}
\usepackage{amssymb}
\usepackage{booktabs}
\usepackage{array}
\usepackage{tabularx}
\usepackage[ruled,vlined,linesnumbered]{algorithm2e}
\DontPrintSemicolon
\SetAlgoNoEnd
\usepackage[a4paper,left=22mm,right=22mm,top=24mm,bottom=24mm,footskip=10mm]{geometry}
\usepackage[section]{placeins}
\newcolumntype{L}[1]{>{\raggedright\arraybackslash}p{#1}}


\journal{Chaos, Solitons and Fractals}

\begin{document}
\begin{frontmatter}

\title{Payoffs and perception mediate environmental feedback in an $N$-player trust game with $Q$-learning}

\author[inst1]{Ruqiang Guo}

\affiliation[inst1]{organization={College of Mathematics and Computer, Guilin Normal College},
            addressline={China},
            city={Guilin},
            postcode={541000},
            state={Guangxi},
            country={China}}

\author[inst2]{Zhaoyi Hu}
\author[inst1]{Fangfang Wang}
\author[inst2]{Linjie Liu\corref{cor}}
\cortext[cor]{Corresponding author}
\ead{linjieliu1992@nwafu.edu.cn}

\affiliation[inst2]{organization={College of Science, Northwest A\&F University},
            addressline={China},
            city={Yangling},
            postcode={712100},
            state={Shaanxi},
            country={China}}

\begin{abstract}
Trust develops through learning, while collective behavior can alter the
environment in which later decisions are made. Reinforcement-learning models
describe adaptation, and eco-evolutionary models describe behavior-environment
feedback, but how an endogenous environment changes trust through material
incentives and perceived states remains unclear. We couple a fixed-role,
two-population hierarchical trust game populated by heterogeneous tabular
$Q$-learning agents to a centered endogenous environment. For the main feedback
comparisons, we initialize the fixed-payoff baseline and feedback branches
from the same learned state to control for prior learning. We then compare
responses when payoff parameters or the observed environmental tier vary. As the
payoff multiplication factor increased, successful trust in the fixed-payoff
game crossed from a low to a higher finite-horizon level, led by risk-seeking
investors. Environmental feedback promoted trust near perception-tier boundaries and
reinforced its decline at the high-return operating point. At the high-return operating point, payoff feedback reinforced
the learning-driven decline in trust. Stronger drive and longer horizons
deepened this negative effect. Compared with a fixed-payoff trust
game, environmental feedback carries past collective outcomes into future
payoffs and observed states. 
\end{abstract}

\end{frontmatter}

\section{Introduction}
\label{sec:Int}

Trust decisions expose individuals to uncertain returns from others' actions.
The trust game introduced by Berg et al.~\cite{Berg1995}
represents this problem through a sequential exchange. An investor decides whether to transfer resources, the transferred amount
is multiplied, and a trustee decides how much to return. The investor risks exploitation, while the trustee can
retain resources or reciprocate. A meta-analysis found that investors
transferred about $50\%$ of their endowment and trustees returned about
$37\%$~\cite{Johnson2011}, departing from the self-interested one-shot
equilibrium. Risk attitudes, time preferences, and prior experience also
shape trust decisions~\cite{Ashraf2006,Fehr2009}. Repeated interaction adds
learning and adaptation, which can change how trust develops over time.
Trust also belongs to the broader study of human cooperation and moral
preferences~\cite{Perc2017Statistical,Szolnoki2020Gradual,Szolnoki2021Averaged,Szolnoki2022Tactical,Capraro2021Moral}.

Evolutionary models of the $N$-player trust game and its replicator
dynamics~\cite{Abbass2016The} examine how reputation, punishment, exclusion,
and regulation sustain trust in structured
populations~\cite{Liu2025inter,Xia2022,Zhou2025,Kumar2020,Guo2023HierarchicalTrust,Liu2023Gov,Sun2022PunishingTrust}.
Related trust-game studies examine diverse investment
patterns~\cite{Shang2023Investment}, conditional investment in repeated
groups~\cite{Liu2022Conditional}, and fixed provider and consumer roles
under different network structures~\cite{Chiong2022Sharing}. These models
typically use imitation, in which agents copy successful strategies.
Reinforcement learning (RL) instead formalizes adaptation through an agent's
own trial-and-error experience~\cite{Watkins1992}. Coupling RL with evolutionary
games produces varied dynamics in cooperation
dilemmas~\cite{Tuyls2006,Zheng2024PGG,PhysRevE.111.014304,Xie2025DoubleQ,Lin2025Coevolution,Lv2025SpatialRL,Yang2025SpatialRL}.
In trust games, $Q$-learning can generate high trust and
trustworthiness~\cite{Zheng2024DecodingTrust}, and reputation-coupled variants
extend to networked and higher-order
populations~\cite{Zhu2025SecondOrder,Hu2026HigherOrder}.


Coevolutionary games couple strategic behavior to changes in the conditions
of interaction~\cite{Perc2010Coevolutionary}.
Eco-evolutionary models also allow collective behavior to change the
environment and hence payoffs, producing oscillations and tipping
points~\cite{Weitz2016,Tilman2020EnvFeedback}. Trust-game models include
margin-driven environmental feedback~\cite{Zhang2025Margin}, loss
assessment~\cite{Liu2024Loss}, and reputation-based nonlinear
investment~\cite{Jiang2026Nonlinear}, within replicator or imitation dynamics.
RL with environmental feedback has been studied chiefly in spatial
public-goods games~\cite{Lv2025SpatialRL,Yang2025SpatialRL}. In a hierarchical
trust game, the environment can affect both the payoff received and the
state on which a learned policy is conditioned. How these two routes shape
trust remains unclear, particularly when the learning agents differ in their
responses to gains and losses. Prospect theory provides a basis for
representing this behavioral heterogeneity~\cite{Kahneman1979,Tversky1992}.

Here, we study a two-population, fixed-role hierarchical trust game with tabular
$Q$-learning, type-dependent rewards, and trustee discount factors.
Collective outcomes drive an environmental state $E_t$, which maps to the
payoff multiplication factor. For the main feedback comparisons, we initialize
the fixed-payoff baseline and feedback branches from the same learned state. Each feedback branch follows its own environmental path.

Environmental feedback promoted trust near perception-tier boundaries but reinforced trust erosion at high-return operating points. At high-return operating points, payoff feedback reinforced the decline in trust driven by the learning process. One-round thresholds
predict the participation ordering, while a discount-factor control shows
how continuation value advances the learned onset.
Together, these results show how past collective outcomes feed back through
future payoffs and observed states. Perceived conditions can promote trust
near a tier boundary, whereas payoff feedback can reinforce trust erosion
at high returns.

Section~\ref{sec:Mod} defines the model, centered feedback, and simulation
protocol. Section~\ref{sec:Res} presents the baseline dynamics and paired
feedback responses. Section~\ref{sec:Mech} connects the mechanistic analysis
with robustness checks, including type-specific behavior, threshold benchmarks,
and finite-window outcome-pair dependence. Section~\ref{sec:Con}
discusses the findings.

\section{Model and methods}
\label{sec:Mod}

\subsection{Populations and the trust game}
\label{sec:Protocol}

\begin{figure*}[!t]
	\centering
	\includegraphics[width=1\textwidth]{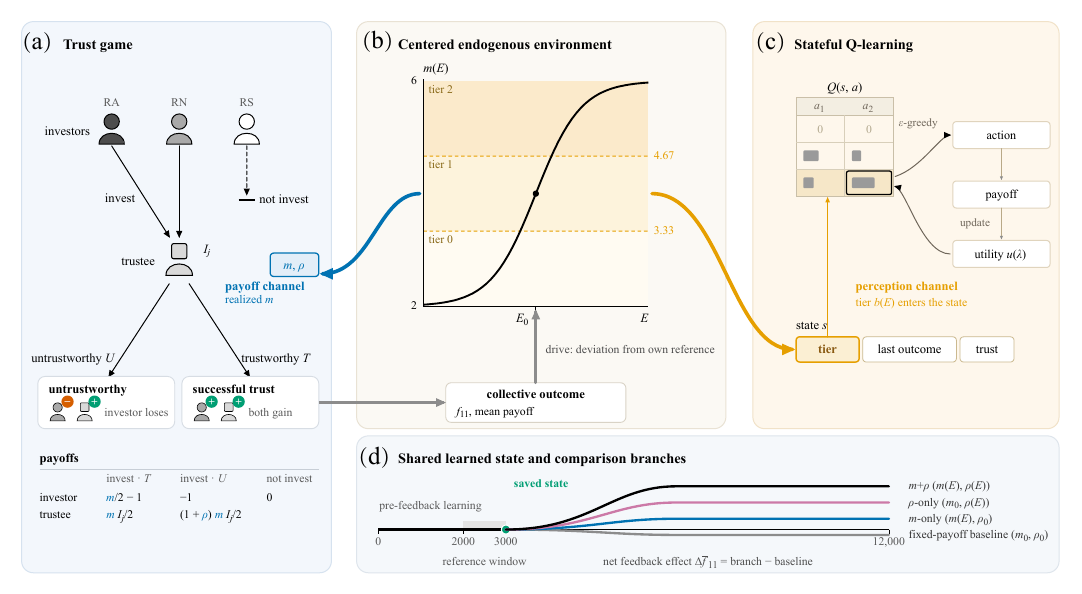}
  \caption{Model and paired comparisons with centered environmental feedback. (a) Investors choose whether
  to invest, matched trustees choose the trustworthy ($T$) or untrustworthy
  ($U$) action. (b) Collective outcomes update the environment, which enters
  realized payoffs through $(m_t,\rho_t)$ and decision states through the
  public tier $b_E(t)$. (c) Agents use tabular $\varepsilon$-greedy
  $Q$-learning. (d) The learned state saved at round 3000 initializes the
  fixed-payoff baseline and the $m$-only, $\rho$-only and $m{+}\rho$ branches. Each branch
  uses the seed's reference over $[2000,3000)$, feedback branches evolve
  their own environments.}
  \label{fig:model-schematic}
\end{figure*}

We model investors $\mathcal{V}$ ($|\mathcal{V}|=N_I$) and
trustees $\mathcal{T}$ ($|\mathcal{T}|=N_T$) with fixed roles and
discrete-time interactions. Each round, investor $i$ is matched uniformly
at random to trustee $j(i,t)$, allowing many-to-one matching. The interaction
follows the sequential trust-game
protocol~\cite{Berg1995,Abbass2016The}. Investor $i$ chooses
$a_i(t)\in A_I=\{0,1\}$, corresponding to not investing or investing one unit
of capital. Trustee $j$ then chooses the trustworthy ($T$) or untrustworthy
($U$) action~\cite{Abbass2016The}. These choices determine the investor payoff,
\begin{equation}
  r_i(t)=\begin{cases} m(t)/2-1, & a_i(t)=1,\ a_j(t)=T,\\
  -1, & a_i(t)=1,\ a_j(t)=U,\\ 0, & a_i(t)=0,\end{cases}
  \label{eq:investor-payoff}
\end{equation}
and, with $I_j(t)=\sum_{i\in\mathcal{M}_j(t)}a_i(t)$ the number of matched
investors who invest and $\rho(t)\in(0,1)$ the defect ratio~\cite{8094275},
\begin{equation}
  w_j(t)=\begin{cases}\dfrac{m(t)}{2}\,I_j(t), & a_j(t)=T,\\[0.5em]
  \dfrac{(1+\rho(t))\,m(t)}{2}\,I_j(t), & a_j(t)=U.\end{cases}
  \label{eq:trustee-payoff}
\end{equation}
The defect ratio raises the untrustworthy trustee's return by
$\rho(t)m(t)I_j(t)/2$. Investor losses are defined separately in
Eq.~\eqref{eq:investor-payoff}, so changing $\rho$ changes total surplus.
Figure~\ref{fig:model-schematic} summarizes one interaction round and the
paired comparisons initialized from the same learned state at round 3000.

\subsection{Q-learning with behavioral heterogeneity}
\label{sec:QLearning}

Both populations use tabular $Q$-learning with $\varepsilon$-greedy
exploration and type-dependent reward functions. Investor types are labeled
risk-averse, risk-neutral, and risk-seeking,
$\theta\in\{\mathrm{RA},\mathrm{RN},\mathrm{RS}\}$.
The loss aversion coefficient sets the slope for negative payoffs, with
$\lambda_{\mathrm{RA}}>1=\lambda_{\mathrm{RN}}>\lambda_{\mathrm{RS}}$
in the prospect-theory-inspired value function~\cite{Kahneman1979}
\begin{equation}
  u_\theta(r)=\begin{cases} r, & r\ge 0,\\ \lambda_\theta\, r, & r<0.\end{cases}
  \label{eq:utility-investor}
\end{equation}
We choose $\lambda_{\mathrm{RA,RN,RS}}=(2.0,1.0,0.5)$ as contrasting
model values for loss weights above, equal to, and below the unit gain weight.
We assign trustees to long-sighted ($L$) and short-sighted
($S$) types whose utilities add a reputation term,
\begin{equation}
  u_\phi(w_j,\Delta R_j)=w_j+\lambda_\phi\,\Delta R_j,
  \qquad \lambda_{L}>\lambda_{S},\quad
  \gamma_{L}>\gamma_{S},
  \label{eq:utility-trustee}
\end{equation}
with $\Delta R_j:=R_j(t{+}1)-R_j(t)$. Investor type enters only through the
loss aversion coefficient $\lambda_\theta$ in Eq.~\eqref{eq:utility-investor}, whereas
trustee type enters through both the reputation weight $\lambda_\phi$ in
Eq.~\eqref{eq:utility-trustee} and the discount factor $\gamma_\phi$ in
Eq.~\eqref{eq:QL-trustee}. Both actions remain available to every agent under $\varepsilon$-greedy
selection.

$Q$-values follow standard temporal-difference updates,
\begin{align}
  Q_i(s_i,a_i) &\leftarrow (1-\alpha_Q)Q_i(s_i,a_i)
    +\alpha_Q\bigl[u_{\theta(i)}(r_i)+\gamma\max_{a'}Q_i(s_i',a')\bigr],
  \label{eq:QL-investor}\\
  Q_j(s_j,a_j) &\leftarrow (1-\alpha_Q)Q_j(s_j,a_j)
    +\alpha_Q\bigl[u_{\phi(j)}(w_j,\Delta R_j)+\gamma_\phi\max_{a'}Q_j(s_j',a')\bigr].
  \label{eq:QL-trustee}
\end{align}
Reputations and pairwise trust weights evolve by exponential smoothing,
\begin{equation}
  R_j(t{+}1)=(1-\eta_R)R_j(t)+\eta_R\chi_j(t),\qquad
  T_{ij}(t{+}1)=(1-\eta_T)T_{ij}(t)+\eta_T\chi_{ij}(t),
  \label{eq:rep-trust-update}
\end{equation}
with $\chi_j=\mathbf{1}[I_j>0]\,\mathbf{1}[a_j=T]$ and
$\chi_{ij}=\mathbf{1}[a_j=T]$ for matched invested pairs.

Investor states encode the last outcome, the matched trustee's discretized
trust level, and the public environment tier
($|\mathcal{S}_I|=4\times3\times3=36$).
Trustee states encode own reputation, inflow, last action, average received
trust, and the environment tier ($|\mathcal{S}_T|=162$).
All $Q$-tables start at zero, with uniform tie-breaking and exploration.
Each binary action therefore has probability $1/2$ when an unupdated state
row is first visited. Complete state definitions and binning functions are
given in \ref{app:state-space}.

\subsection{Centered environment and the two channels}
\label{sec:EnvFeedback}

To represent feedback between collective behavior and its
payoff environment~\cite{Weitz2016,Tilman2020EnvFeedback}, we define a
macroscopic state $E_t\in\mathbb{R}$ and a latent payoff-multiplication signal,
\begin{equation}
  \begin{aligned}
    m_{\rm signal}(E_t)&=m_{\min}+(m_{\max}-m_{\min})\,
    \sigma\!\bigl(kE_t\bigr),\\
    \sigma(x)&=\frac{1}{1+e^{-x}}.
  \end{aligned}
  \label{eq:env-to-m}
\end{equation}
The environment affects agents through two channels. The \emph{payoff
channel} is the realized pair $(m_t,\rho_t)$ entering
Eqs.~\eqref{eq:investor-payoff}--\eqref{eq:trustee-payoff}.
The \emph{perception channel} is the public environment tier
$b_E(t)=b_3\bigl[(m_{\rm signal}(E_t)-m_{\min})/(m_{\max}-m_{\min})\bigr]$,
a component of every agent's decision state.
Here $b_3$ is the three-level map defined in \ref{app:state-space}.
The tier boundaries lie at
$m_{\rm signal}=m_{\min}+(m_{\max}-m_{\min})/3=10/3\simeq3.333$ and
$m_{\min}+2(m_{\max}-m_{\min})/3=14/3\simeq4.667$.
Entering a previously unvisited tier activates $Q$ rows with no earlier
updates; returning to a visited tier can reuse updated rows.
The main payoff-channel branches apply this map to their own latent
signals while differing in the payoff parameters they realize.
Supplementary perception controls can instead hold the observed tier fixed.

For the dynamic-$\rho$ controls, the defect-ratio tier is centered on the
nominal environment,
\begin{equation}
  \begin{aligned}
    b_\rho(E_t)&=b_3\!\left[\sigma\!\left(k(E_t-E_0)\right)\right],\\
    \rho_{\rm dyn}(E_t)&=\rho_{b_\rho(E_t)},\\
    (\rho_{L},\rho_{M},\rho_{H})
      &=(0.5,1,1.5)\times\rho_0 ,
  \end{aligned}
  \label{eq:rho-tiers}
\end{equation}
so that $\rho_{\rm dyn}(E_0)=\rho_0$ and, because the tiers of
Eq.~\eqref{eq:bin3} are half-open, the tier leaves the middle band only when
$E_t-E_0<-\ln2/k$ or $E_t-E_0\ge\ln2/k$ ($|E_t-E_0|\simeq0.173$). We
distinguish three payoff-channel
interventions:
\begin{equation}
 (m_t,\rho_t)=
 \begin{cases}
  (m_{\rm signal}(E_t),\rho_0), & m\text{-only (main)},\\
  (m_0,\rho_{\rm dyn}(E_t)), & \rho\text{-only},\\
  (m_{\rm signal}(E_t),\rho_{\rm dyn}(E_t)), & m{+}\rho.
 \end{cases}
 \label{eq:payoff-channels}
\end{equation}
The $\rho$-only branch provides a perception control when its
$\rho$-tier stays at the nominal value. Under this condition, both payoff
parameters remain fixed, and the imposed change from the static baseline is
the dynamic perceived state. Each feedback branch evolves its own $E_t$
and perceived tier sequence. Where the $\rho$-tier changes, the
$\rho$-only branch also changes realized payoff parameters.
Hereafter, environment-feedback Q-learning (EF-Q) denotes the $m$-only main branch unless a channel is named
explicitly.

The environment is driven by the volume-based cooperation rate
$C_t:=f_{11}(t)$ (the population share of successful-trust interactions,
Section~\ref{sec:SimSetup}) and the population-mean investor payoff
$\bar r_{I,t}$ (averaged over \emph{all} investors, with non-investors
contributing zero). Define
\begin{equation}
 \begin{aligned}
  D_t&=\beta_C(C_t-C_{\rm ref})
       +\beta_r(\bar r_{I,t}-r_{\rm ref}),\\
  E_{t+1}&=E_0+(1-\nu)(E_t-E_0)+\nu D_t+\xi_t,
 \end{aligned}
  \label{eq:env-update}
\end{equation}
where $\nu$ is the environmental update rate and $\xi_t$ is optional Gaussian
noise. The fixed anchor $E_0$ satisfies $m_{\rm signal}(E_0)=m_0$.
At the reference observables, $D_t=0$, so a noiseless environment initialized
at $E_0$ remains there. For constant $D$ and zero noise, the environment
approaches $E_0+D$ with relaxation time $-1/\log(1-\nu)\simeq1/\nu$.
The change in any one round is
\[
 E_{t+1}-E_t=\nu\bigl[D_t-(E_t-E_0)\bigr]+\xi_t.
\]
Thus a positive drive can coexist with a falling environmental level.
At fixed $E_t$ and the other drive input, increasing either observable raises $E_{t+1}$
and the next payoff-multiplication signal because both weights are positive.
The loop-gain analysis below describes how behavioral responses to $m$
modify this environmental relaxation.

For each nominal point and seed, we first run rounds 0--2999 with
$(m_t,\rho_t)=(m_0,\rho_0)$ and $E_t=E_0$. This pre-feedback period visits only
the nominal environment tier, leaving the $Q$ rows for the other two tiers
unupdated. We define the seed-specific reference values
$(C_{\rm ref},r_{\rm ref})$ as the time averages over rounds $[2000,3000)$.
At round 3000, we save the complete learned state, memory variables, aggregate
counters, and named random-number streams. We then clone this state to
initialize the fixed-payoff baseline and the three feedback branches in
Eq.~\eqref{eq:payoff-channels}. Because every branch starts from the same
seed-specific learning history, their within-seed differences measure the
finite-horizon effect of the imposed feedback, with the reference values held
fixed.

\subsection{Simulation settings and statistical methods}
\label{sec:SimSetup}

After round 3000, we hold $(m_t,\rho_t)=(m_0,\rho_0)$ fixed to define the
fixed-payoff baseline. In the main EF-Q branch, we update $E_t$ through
Eq.~\eqref{eq:env-update}, allow $m_t$ to vary, and keep $\rho_t=\rho_0$.
The $\rho$-only and $m{+}\rho$ branches isolate the additional channel changes
defined above.
The nominal anchor
$E_0(m_0)=\sigma^{-1}[(m_0-m_{\min})/(m_{\max}-m_{\min})]/k$
is finite only for $m_{\min}<m_0<m_{\max}$.
The $20\times20$ feedback grid uses cell centers, with
$m_0=2.1,2.3,\ldots,5.9$ and
$\rho_0=0.06125,0.08375,\ldots,0.48875$ (step $0.0225$).
All branch contrasts are paired within the same point--seed checkpoint.
Algorithm~\ref{alg:environment-feedback} gives the simulation and paired-branch protocol.

\begin{algorithm}[htbp]
\small
\caption{Environment-feedback $Q$-learning in the trust game}
\label{alg:environment-feedback}
\KwIn{Nominal parameters $(m_0,\rho_0)$, random seed, and model parameters in Table~\ref{tab:parameters}.}
\KwOut{Successful-trust trajectories, window statistics, and paired feedback effects.}
Assign investor and trustee types; set $Q_i=Q_j=0$, $R_j=T_{ij}=0.5$, and $E=E_0$\;
Set $X_i(-1)=00$ and $a_j(-1)=U$; start with the static baseline\;
\For{$t=0,\ldots,T_{\max}-1$}{
  \If{$t=3000$}{
    Compute $(C_{\rm ref},r_{\rm ref})$ from rounds $[2000,3000)$ and hold them fixed\;
    Clone the complete state, previous transitions, and random-number streams into the baseline, $m$-only, $\rho$-only, and $m{+}\rho$ branches\;
  }
  \ForEach{current branch}{
    Set $(m_t,\rho_t)=(m_0,\rho_0)$ in the baseline; otherwise use Eq.~\eqref{eq:payoff-channels}.
    Compute $b_E(t)$ using Eq.~\eqref{eq:environment-bin}\;
    Match each investor $i$ uniformly to a trustee $j(i,t)$\;
    Construct all investor states $s_i(t)$ using Eq.~\eqref{eq:investor-state}\;
    \lIf{$t>0$}{update each investor's previous transition by Eq.~\eqref{eq:QL-investor}, with $s_i'=s_i(t)$}
    Select each $a_i(t)$ by $\varepsilon$-greedy exploration with uniform tie-breaking\;
    Aggregate $I_j(t)$ and construct all trustee states $s_j(t)$
    (Eqs.~\eqref{eq:normalized-inflow} and \eqref{eq:trustee-state})\;
    \lIf{$t>0$}{update each trustee's previous transition by Eq.~\eqref{eq:QL-trustee}, with $s_j'=s_j(t)$}
    Select each $a_j(t)$ by the same exploration rule, including trustees with $I_j(t)=0$\;
    Compute payoffs $r_i(t)$ and $w_j(t)$ using Eqs.~\eqref{eq:investor-payoff}--\eqref{eq:trustee-payoff}\;
    Update $R_j$, $T_{ij}$, and $\Delta R_j$ using Eqs.~\eqref{eq:rep-update-app}--\eqref{eq:trust-update-app}\;
    Compute subjective rewards using Eqs.~\eqref{eq:utility-investor}--\eqref{eq:utility-trustee};
    retain each agent's current state, action, and reward for the next $Q$ update\;
    Record $X_i(t)$, $C_t$, $\bar r_{I,t}$, and $f_{11}(t)$\;
    \eIf{this is a feedback branch}{
      Update $E_{t+1}$ using Eq.~\eqref{eq:env-update}\;
    }{
      Keep $E_{t+1}=E_0$\;
    }
  }
}
Compute window statistics and within-seed feedback-minus-baseline contrasts\;
\end{algorithm}

We define the outcome state of each pairwise interaction as
$s^{\text{out}}=(v_i,h_j)$, where
$s^{\text{out}}\in\{00,01,10,11\}$, with
$v_i=a_i$ and $h_j=\mathbf{1}[a_j=T]$; $f_{ab}(t)$ denotes the
population share of state $ab$. We call $f_{11}(t)$ the successful trust
frequency. With $\mathcal W=\{3000,\ldots,11999\}$ the
specified analysis window and $S$ independent seeds, we define
\begin{equation}
  \bar f_{ab}:=\frac{1}{S}\sum_{s=1}^{S}\frac{1}{|\mathcal W|}
  \sum_{t\in\mathcal W}f_{ab}^{(s)}(t).
  \label{eq:finite-window-mean}
\end{equation}
The primary estimand is the net feedback effect
$\Delta\bar f_{11}^{M}:=\bar f_{11}^{m\text{-only}}-
\bar f_{11}^{\text{base}}$, computed as the within-seed paired change between
the $m$-only branch and the static baseline; channel-control contrasts are
defined analogously.
We also record the baseline learning drift and realized environmental
excursion for each seed. Baseline learning drift is the change in successful
trust generated by continued learning under fixed payoff parameters,
$\delta^{(s)}:=\langle f_{11}^{(s)}\rangle_{\mathcal W}-\langle
f_{11}^{(s)}\rangle_{[2000,3000)}$. Economically, it measures endogenous
erosion or recovery of trust under unchanged material incentives and separates
ongoing adaptation from the additional effect of environmental feedback.
The realized excursion is the window
mean of the realized payoff multiplication factor minus its nominal value,
$\Delta\bar m^{(s)}:=\langle m^{(s)}_t\rangle_{\mathcal W}-m_0$.

At each $\rho$, we fit the fixed-horizon response $\bar f_{11}(m)$ with
\begin{equation}
  f(m)=\frac{A}{1+\exp[-(m-m^*)/w]}+c.
  \label{eq:crossover-fit}
\end{equation}
We report the fitted midpoint $m^*$ and crossover band $m^*\pm w$.
The cross-seed variance peak summarizes heterogeneity over the same window.

Over $\mathcal W$, we define $P_o(s,s')$ as the empirical joint frequency of
consecutive outcome-state pairs and $P_e(s,s'):=\pi(s)\pi(s')$ as the shuffled
independence benchmark built from the pooled same-window state frequencies.
Following the cellwise normalization of Li et al.~\cite{PhysRevE.111.014304},
we compute the normalized excess transition index (NETI),
\begin{equation}
  \tilde{\kappa}(s,s'):=\frac{P_o(s,s')-P_e(s,s')}{1-P_e(s,s')}
  \label{eq:kappa}
\end{equation}
which measures excess pair frequency with a chance correction analogous to
Cohen's~$\kappa$~\cite{Cohen1960}. A positive diagonal entry $\tilde\kappa(s,s)$ indicates
self-persistence relative to the shuffled same-window benchmark.
A positive off-diagonal entry identifies an over-represented directed
outcome pair. Pooling across agents mixes within-agent sequence dependence,
heterogeneity in individual action rates, and slow drift within
$\mathcal W$.
NETI summarizes these pooled dependencies. Its relation to conditional
transition probabilities and its feasible bounds are derived in
\ref{app:kappa-details}.

In the main runs, we simulate 240 investors and 120 trustees for 12,000
rounds, with a burn-in cutoff at round 3,000. Table~\ref{tab:parameters}
lists the baseline model and simulation parameters; control-specific changes
accompany each experiment.
The main parameter and control runs use 20 seeds, the A--D
comparisons use 50. We use 10 seeds to examine longer-term baseline dynamics
(Fig.~\ref{fig:convergence-diagnostic}).
With learning and exploration rates held constant, we analyze the specified
window $\mathcal W$.
We extended the simulations to 36,000 rounds and found continued trust
decline at two parameter points and a late recovery at the third (\ref{app:supp-experiments}).
Unless noted, error bands are two-sided pointwise 95\%
Student-$t$ confidence intervals across the stated seed units.
Bootstrap intervals are identified separately, and seed-level Pearson
correlations describe association across matched seeds.
The per-figure run specifications are listed in \ref{app:parameters}.

\begin{table}[!htbp]
  \centering\footnotesize
  \caption{Baseline model and simulation parameters. Control-specific
  changes are stated with each experiment.}
  \label{tab:parameters}
  \begin{tabularx}{\linewidth}{@{}L{0.17\linewidth}L{0.22\linewidth}L{0.23\linewidth}>{\raggedright\arraybackslash}X@{}}
    \toprule
    Group & Symbol & Value & Meaning \\
    \midrule
    Population & $N_I,\ N_T$ & 240, 120 & investors, trustees (main runs) \\
    Learning & $\alpha_Q$ & 0.1 & learning rate (both roles) \\
     & $\gamma$ & 0.9 & investor discount factor \\
     & $\gamma_{L},\gamma_{S}$ & 0.95, 0.80 & trustee discount factors \\
     & $\varepsilon$ & 0.05 & exploration rate (constant) \\
    Types & $\lambda_{\mathrm{RA,RN,RS}}$ & (2.0, 1.0, 0.5) & loss aversion coefficients \\
     & $\lambda_{L,S}$ & (1.0, 0.1) & trustee reputation weights \\
     & fractions & $1/3$ each; $1/2$ each & type mixtures \\
    Reputation & $\eta_R,\ \eta_T$ & 0.05, 0.05 & reputation and trust smoothing \\
     & $R_0,\ T_0$ & 0.5, 0.5 & initial values; $Q\equiv 0$ \\
    Discretization & $\kappa_1,\kappa_2$ & $1/3,\ 2/3$ & three-tier binning; tier edges at $m=3.333,\ 4.667$ \\
    Environment & $\nu$ & 0.03 & adjustment speed \\
     & $(\beta_C,\beta_r)$ & (1.4, 1.2) & drive weights \\
     & $k$ & 4.0 & sigmoid slope \\
     & $(m_{\min},m_{\max})$ & (2, 6) & $m$ range \\
     & $\rho$ tiers & $(0.5,1,1.5)\times\rho_0$ & centered relative tiers \\
     & $\sigma_\xi$ & 0 (main); $\{0.002,0.005\}$ & noise (robustness) \\
     & $E_0(m)$ & $\sigma^{-1}\!\bigl(\tfrac{m-2}{4}\bigr)/k$ & nominal environmental anchor \\
    Runs & $T_{\max}$, burn-in cutoff & 12000, 3000 & fixed-horizon preset \\
     & seeds & 20 main; 50 A--D; 10 S1 & per condition \\
    \bottomrule
  \end{tabularx}
\end{table}

\section{Results}
\label{sec:Res}

\begin{figure}[!htbp]
	\centering
\includegraphics[width=1\textwidth]{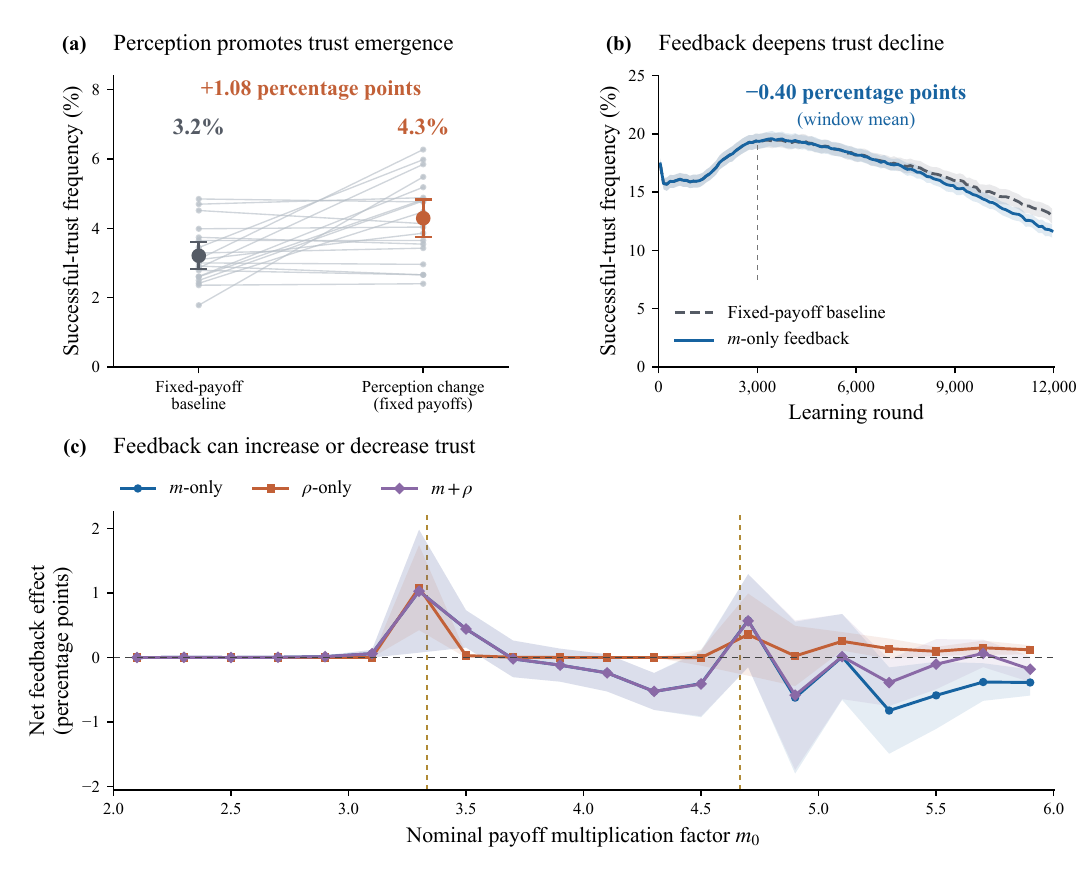}
  \caption{Environmental feedback promotes trust emergence or deepens trust
  decline under different conditions. All panels use $\rho_0=0.2075$.
  Net feedback effects are within-seed differences in mean successful-trust
  frequency from the fixed-payoff baseline. (a) At $m_0=3.3$, the $\rho$-only
  branch changes the perceived environmental state while both realized payoff
  parameters remain fixed. Thin lines connect paired outcomes; large points
  and error bars show means and 95\% Student-$t$ intervals. The paired increase
  is 1.08 percentage points (95\% interval: 0.42--1.74). (b) Baseline and
  $m$-only trajectories at D ($m_0=5.7$), shown as 100-round means. The vertical
  dashed line marks feedback activation at round 3,000. The analysis-window
  paired change is $-0.40$ percentage points (95\% interval: $-0.56$ to $-0.24$).
  (c) Net feedback effects across nominal payoff multiplication factors.
  The $m$-only, $\rho$-only and joint branches allow $m$, $\rho$, or both payoff
  parameters to respond to each branch's environment, which also determines
  the perceived state. The horizontal dashed line denotes zero effect; gold
  vertical lines mark the perception-tier boundaries. Panels (a,c) use 20
  paired seeds; (b) uses 50 additional paired seeds. Shading in (b,c) represents
  pointwise 95\% Student-$t$ intervals. Window means use rounds 3,000--11,999.
  Perception promotes trust emergence at fixed payoff parameters in (a),
  whereas feedback deepens the decline in (b).}
  \label{fig:main-result-overview}
\end{figure}

Environmental feedback changed trust in opposite directions under different
conditions (Fig.~\ref{fig:main-result-overview}). We compared branches that
continued $Q$-learning from the same learned state at round 3,000; the static
baseline kept the payoff parameters and environmental state fixed. We measured
trust by the successful-trust frequency $f_{11}$, the share of interactions in
which the investor invested and the trustee acted trustworthily. The net
feedback effect was
\[
  \Delta\bar f_{11}=\bar f_{11}^{\mathrm{feedback}}-
  \bar f_{11}^{\mathrm{base}}.
\]
The bars denote means over rounds 3,000--11,999 and paired seeds. Positive and
negative effects indicate higher and lower trust than the baseline, respectively.

At $\rho_0=0.2075$ and $m_0=3.3$, changing the perceived environmental state
in the $\rho$-only branch increased successful-trust frequency from 3.2\% to
4.3\%, a paired increase of 1.08 percentage points
(Fig.~\ref{fig:main-result-overview}a). Both realized payoff parameters
remained fixed: perception alone promoted the emergence of trust. At point D
($m_0=5.7$), $m$-only feedback instead deepened the decline in trust, reducing
the analysis-window mean by 0.40 percentage points relative to the baseline
(Fig.~\ref{fig:main-result-overview}b). Across the three feedback branches,
the direction and magnitude of the response varied with the nominal payoff
multiplication factor (Fig.~\ref{fig:main-result-overview}c). We next examine
how the payoff and perception channels, state representation, and learning
history produce these responses.

\FloatBarrier
\subsection{Baseline finite-horizon crossover}
\label{sec:BaselinePhase}
\begin{figure}[!htbp]
  	\centering
  \includegraphics[width=1\textwidth]{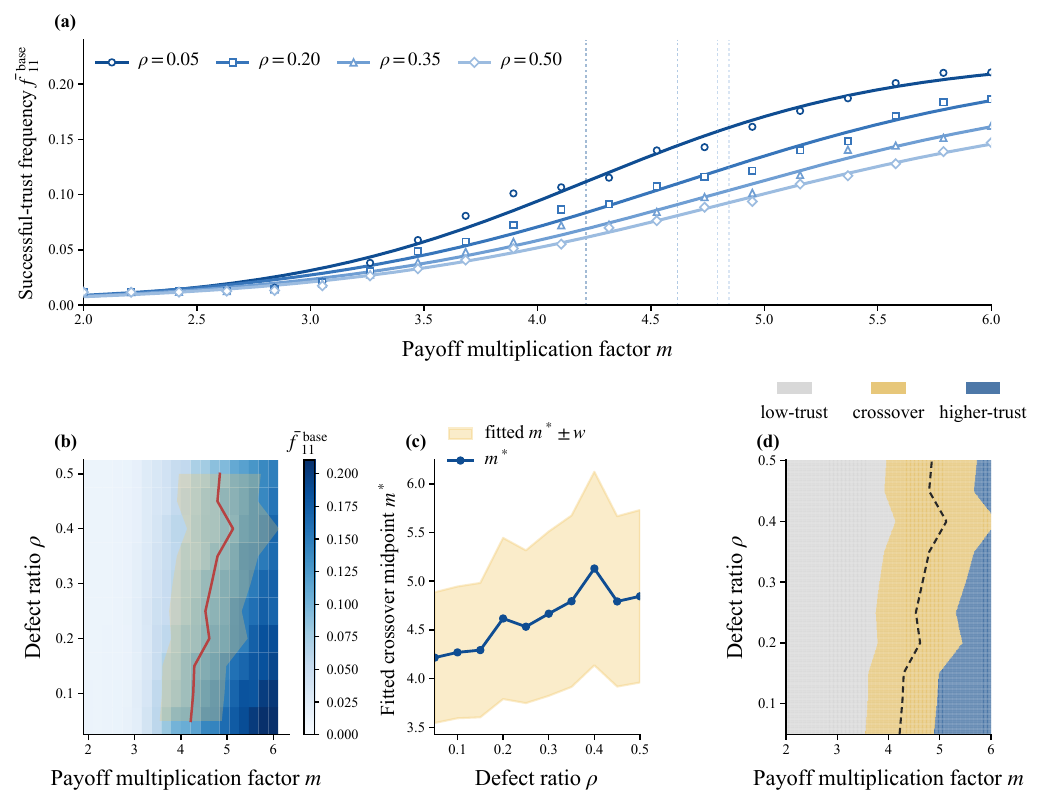}
  \caption{Baseline behavior over the fixed 12,000-round horizon. (a)
  Baseline grid means $\bar f_{11}(m)$ at selected $\rho$ (markers) and
  four-parameter logistic fits (lines). (b) Finite-horizon map
  $\bar f_{11}^{\mathrm{base}}(m,\rho)$ with fitted midpoints and the
  $m^*\pm w$ band. (c) Fitted midpoint $m^*(\rho)$ with shading spanning
  $m^*\pm w$. (d) Fitted crossover band at $T=12{,}000$, separating
  the low-trust, crossover, and higher-trust ranges.
  Each grid value is the mean of 20 seeds over rounds 3,000--11,999.}
  \label{fig:baseline-phase-panel}
\end{figure}

Over the 12,000-round horizon, baseline successful trust increased with $m$ from a
low-trust to a higher-trust range
(Fig.~\ref{fig:baseline-phase-panel}a,b).
Logistic fits to the ten sampled $\rho$ slices gave $R^2=0.990$--$0.997$,
with fitted midpoints $m^*=4.22$--$5.13$ and widths $w=0.67$--$0.99$.
Midpoints generally shifted toward larger $m$ as the defect ratio increased,
with departures from monotonicity between adjacent slices.
The fitted band defines the three ranges in
Fig.~\ref{fig:baseline-phase-panel}d.
Successful trust remained a minority outcome even in the higher-trust
range ($\bar f_{11}\le0.21$), providing the baseline scale for the
feedback comparisons.

\subsection{Baseline learning drift across the crossover}
\label{sec:SlowDynamics}
\begin{figure}[!htbp]
  \centering
    \includegraphics[width=1\textwidth]{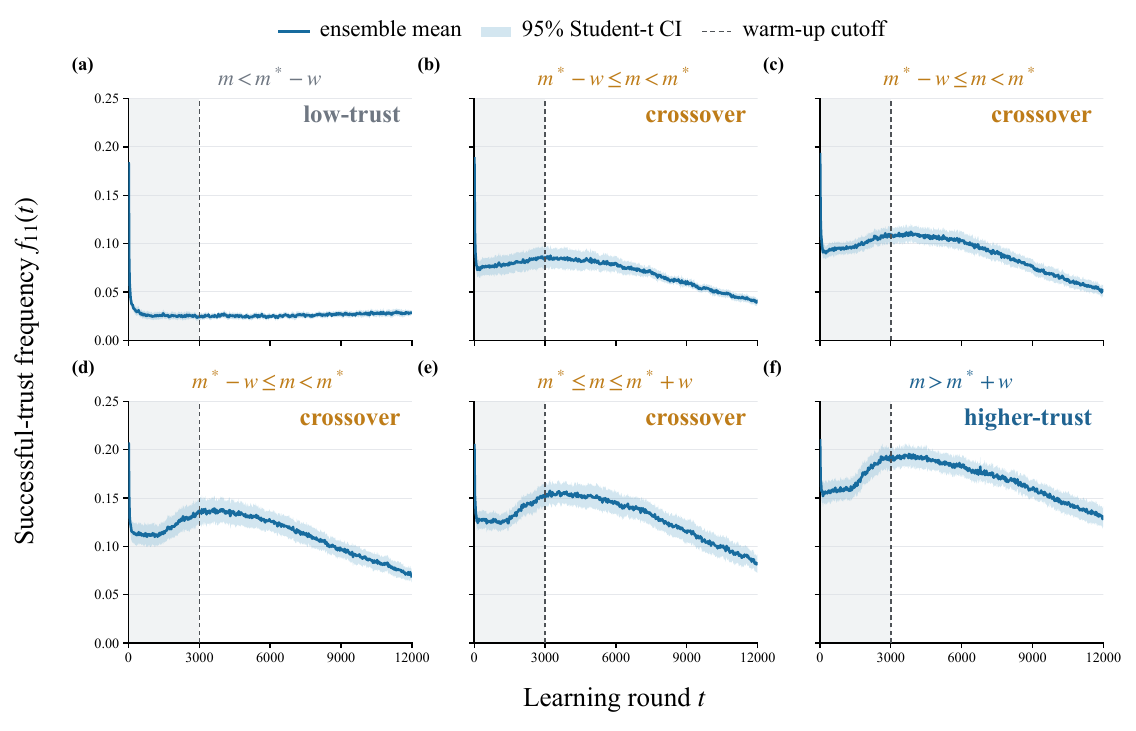}
  \caption{Baseline successful-trust trajectories across six payoff multiplication
  factors at $\rho=0.2075$.
  Panels (a--f) show $f_{11}(t)$ at $m=3.20$, 3.80, 4.20, 4.60, 5.00,
  and 5.60, respectively, on common axes.
  Regime labels use the fitted midpoint $m^*=4.62$ and width $w=0.83$.
  The normalized positions $(m-m^*)/w$ in (a--f) are approximately
  $-1.71$, $-0.99$, $-0.50$, $-0.02$, $+0.46$, and $+1.19$.
  Within-seed 20-round block averages are summarized by the ensemble mean
  and two-sided 95\% Student-$t$ confidence interval across 20 independent
  seeds.
  Shading and the dashed line mark the specified burn-in interval and
  cutoff at round 3,000.
  For $m\ge3.8$, ensemble means peak within about 1,000 rounds of the
  cutoff and then decline over the remaining $T=12{,}000$ horizon.}
  \label{fig:timeseries-parameter-grid}
\end{figure}

Baseline successful-trust trajectories continued to evolve after the
burn-in cutoff (Fig.~\ref{fig:timeseries-parameter-grid}).
At $m=3.2$, the post-cutoff block mean remained near an
exploration-dominated low-trust level, changing from 0.025 at the cutoff to
0.028 in the final block.
Across $m=3.8$--5.6, ensemble means peaked between rounds 3,150 and 3,980
at 0.086--0.194, then declined to 0.040--0.131 in the final block.
The baseline learning drift $\delta$ defined in Section~\ref{sec:SimSetup} was
$+0.001\pm0.001$ at $m=3.2$ and $-0.017$ to $-0.020$ at $m=3.8$--5.6.
All intervals in the latter range excluded zero.
The reference window $[2000,3000)$ lay near the peaks of these ensemble
means, which remained below their reference averages for much of the
analysis window.
Equation~\eqref{eq:env-update} uses each seed's deviation from its own
reference; the plotted curves show ensemble means.

\subsection{Closing the loop: two channels in the response landscape}
\label{sec:ClosingLoop}
\begin{figure}[!htbp]
  \centering
      \includegraphics[width=1\textwidth]{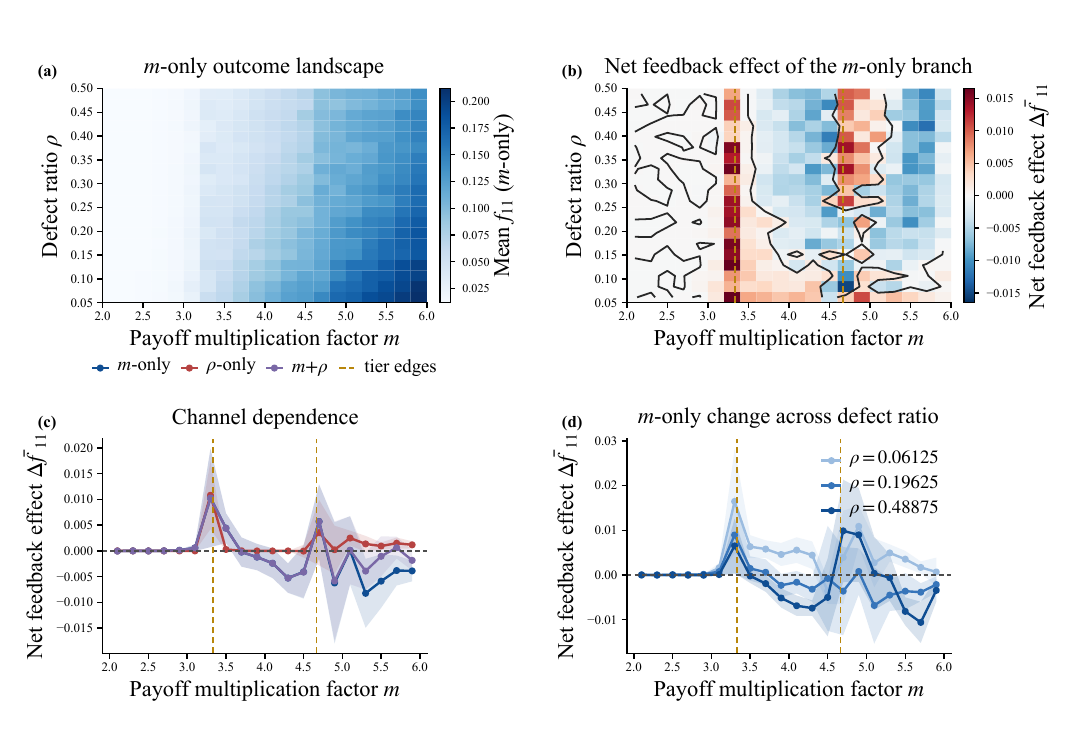}
  \caption{Paired response landscape and payoff-channel controls. (a) Mean
  successful trust frequency $\bar f_{11}$ for the $m$-only branch on the
  $20\times20$ $(m,\rho)$ grid. (b) Within-seed $m$-only-minus-baseline change
  on the same grid; black contours mark zero mean change. Dashed gold lines
  in (b--d) mark the perception-tier edges at $m=3.333$ and $4.667$.
  (c) $m$-only, $\rho$-only, and joint-channel changes along the fixed
  $\rho_0=0.2075$ slice. (d) $m$-only changes along three sampled $\rho$ profiles.
  Grid values in (a,b) are means over 20 paired seeds; lines and bands in
  (c,d) show means and pointwise 95\% Student-$t$ confidence intervals
  across the same 20 matched seeds.}
  \label{fig:feedback-landscape}
\end{figure}

Feedback produced small, parameter-dependent changes in successful trust
(Fig.~\ref{fig:feedback-landscape}). Across the 400 grid points, the paired
$m$-only change in $\bar f_{11}$ ranged from $-0.015$ to $+0.017$ with a
median absolute value of $0.002$; the pointwise intervals lay
above zero at 49 points, below zero at 96, and included zero at 255.
At the sampled $m\le3.1$ points, the population remained at an
exploration-dominated low-trust level.
The maximum absolute paired mean in this range was approximately $0.0021$.
Two narrow regions of positive mean response extended across sampled
$\rho$ values at $m=3.3$ and $m=4.7$.
These columns adjoin the perception-tier edges at $3.333$ and $4.667$.
They contained 21 of the 49 intervals above zero and none of the 96
intervals below zero.
Including neighboring columns at $3.5$ and $4.9$ increased the first count
to 31; at $m=3.3$, 16 of 20 intervals lay above zero.
Over the interior columns $m\ge3.7$, excluding these boundary columns and
their neighbors, the mean change was $-0.003$.
Of these 200 cells, 93 intervals lay below zero and 11 above.

\begin{table}[!htbp]
\centering\small
\caption{Channel-control comparisons along the fixed $\rho_0=0.2075$ slice
(20 matched seeds; means $\pm$ 95\% Student-$t$ half-widths for the
$m$-only and $\rho$-only paired changes $\Delta\bar f_{11}$).
The joint-channel column reports means; its intervals are shown in
Fig.~\ref{fig:feedback-landscape}c. $d_{\rm edge}$ is the distance from the nominal
$m_0$ to the nearest perception-tier edge; $\Delta\bar m$ and $s_m$ are the
window mean excursion and temporal standard deviation of the realized
$m(t)$ in the $m$-only branch. The $\rho$-only branch realizes $m_0$ in every
round; its $\rho$-tier stays in the middle band for $m_0\le4.1$, where the
$m$-only and $m{+}\rho$ full-window successful-trust means agree to numerical
precision within each seed.}
\label{tab:slice-channels}
\setlength{\tabcolsep}{5pt}
\begin{tabular}{@{}rrrrrrr@{}}
\toprule
$m_0$ & $d_{\rm edge}$ & $\Delta\bar f_{11}$ ($m$-only) & $\Delta\bar f_{11}$ ($\rho$-only) & $\Delta\bar f_{11}$ ($m{+}\rho$) & $\Delta\bar m$ & $s_m$ \\
\midrule
3.1 & 0.233 & $+0.0006\pm0.0006$ & $0$ & $+0.0006$ & $+0.006$ & 0.015 \\
3.3 & 0.033 & $+0.0103\pm0.0095$ & $+0.0108\pm0.0066$ & $+0.0103$ & $+0.060$ & 0.052 \\
3.5 & 0.167 & $+0.0044\pm0.0029$ & $+0.0003\pm0.0003$ & $+0.0044$ & $+0.013$ & 0.057 \\
4.1 & 0.567 & $-0.0024\pm0.0029$ & $0$ & $-0.0024$ & $-0.131$ & 0.175 \\
4.5 & 0.167 & $-0.0041\pm0.0052$ & $-0.00004\pm0.0012$ & $-0.0041$ & $-0.198$ & 0.212 \\
4.7 & 0.033 & $+0.0057\pm0.0072$ & $+0.0035\pm0.0064$ & $+0.0057$ & $-0.054$ & 0.117 \\
4.9 & 0.233 & $-0.0062\pm0.0118$ & $+0.0002\pm0.0047$ & $-0.0058$ & $-0.221$ & 0.229 \\
5.3 & 0.633 & $-0.0082\pm0.0067$ & $+0.0014\pm0.0016$ & $-0.0039$ & $-0.216$ & 0.234 \\
5.7 & 1.033 & $-0.0038\pm0.0029$ & $+0.0015\pm0.0011$ & $+0.0006$ & $-0.130$ & 0.139 \\
\bottomrule
\end{tabular}
\end{table}

We fixed $\rho_0=0.2075$ and compared successful trust across the three feedback branches
(Fig.~\ref{fig:feedback-landscape}c and Table~\ref{tab:slice-channels}).
At $m_0=3.3$, near the lower perception-tier boundary, the $m$-only change
was $+0.0103\pm0.0095$, close to the fixed-payoff perception response shown
in Fig.~\ref{fig:main-result-overview}a.
The $m$-only mean excursion was $+0.060$, with temporal standard deviation
$0.052$ and nominal distance $0.033$ to the lower tier boundary.

At $m_0=4.7$, $0.033$ above the second edge, both branch point estimates
were positive: $+0.0035$ for $\rho$-only and $+0.0057$ for $m$-only.
Both 95\% intervals spanned zero at this operating point.
At $m_0=4.5$, $0.167$ below that edge, the mean excursion was directed
away from it ($\Delta\bar m=-0.198$).
The $\rho$-only mean was close to zero and the $m$-only mean was negative;
both 95\% intervals spanned zero.
For $m_0\le4.1$, the $m$-only and $m{+}\rho$ full-window successful-trust
means agreed to numerical precision within each seed, consistent with the
centered $\rho$-tier remaining nominal.

At the listed $m_0\ge4.9$ points, $\rho$-only point estimates were small
and positive ($+0.0002$ to $+0.0015$), opposite to the $m$-only estimates.
These branches allow a declining environment to lower the defect-ratio tier.
Away from the highlighted boundary columns, declining mean payoff multiplication
factors accompanied predominantly negative $m$-only response estimates,
consistent with a contribution from payoff feedback.
The mean decline in the payoff multiplication factor was $0.13$--$0.22$ at the listed
non-boundary points from $m_0=4.1$ to $5.7$.

\subsection{Baseline learning drift and environmental response at four parameter settings}
\label{sec:RepPoints}
\begin{figure}[!htb]
  \centering
        \includegraphics[width=1\textwidth]{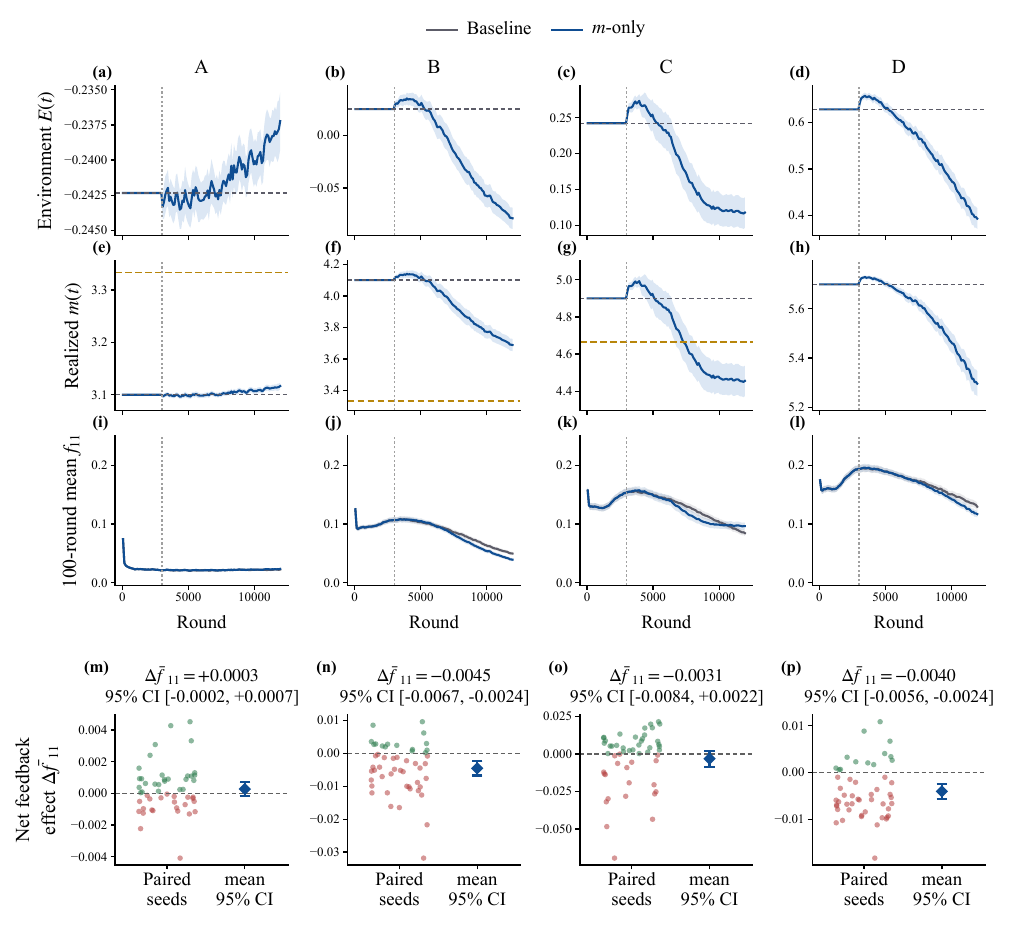}
  \caption{Parameter settings A--D (columns) at $\rho_0=0.2075$
  and $m_0=3.1$, 4.1, 4.9, and 5.7. (a--d) Environment $E(t)$ of the
  $m$-only branch with the nominal $E_0$ (dashed grey). (e--h) Realized
  $m(t)$ with the nominal $m_0$ (dashed grey) and the perception-tier edges
  (dashed gold), summarized as ensemble means. (i--l) Baseline and $m$-only $f_{11}$
  trajectories. (m--p) Within-seed $m$-only-minus-baseline analysis-window
  change for the 50 seeds (green positive, red negative) with the mean and
  95\% Student-$t$ interval. Curves and bands in (a--l) are 100-round means
  and pointwise 95\% Student-$t$ intervals across 50 matched seeds
  (100--149); the vertical dotted line marks the feedback start at round
  3000 and the fixed analysis window is $[3000,12000)$.}
  \label{fig:representative-points}
\end{figure}

We examined four parameter settings using 50 additional random seeds, separate from those used for the parameter grid.
Figure~\ref{fig:representative-points} shows the baseline learning drift and environmental response at these settings.
At A ($m=3.1$), baseline learning drift was $-0.0005\pm0.0006$, the mean
environmental displacement was $\langle E-E_0\rangle=+0.001$, and the paired
change was $+0.0003$ ($95\%$ CI $[-0.0002,+0.0007]$).
At B, C, and D, baseline learning drift was $-0.021$, $-0.021$, and $-0.018$,
respectively, with mean environmental declines of $0.041$, $0.058$, and
$0.076$.
The realized payoff multiplication factor declined by $0.16$, $0.21$, and $0.12$
on average over the window, and by $0.41$, $0.44$, and $0.40$ in the
final bin.
The net feedback effects were $-0.0045$ $[-0.0067,-0.0024]$,
$-0.0031$ $[-0.0084,+0.0022]$, and $-0.0040$ $[-0.0056,-0.0024]$,
respectively (95\% CIs; Table~\ref{tab:selected-neti-ci}).

Environmental excursions were positively associated with baseline learning drift
across seeds. Across the 50 seeds, the Pearson coefficients were $0.76$,
$0.87$, $0.76$, and $0.87$ at A--D, respectively.
The correlations between realized excursions $\Delta\bar m$ and net
feedback effects were $0.65$, $0.91$, $0.94$, and $0.66$.
These are across-seed associations relative to the shared pre-feedback
reference.

At C, inside the fitted crossover band, the excursion--effect correlation
was strongest, while the interval for the mean feedback change included zero.
The three seeds whose environments rose had paired changes from $+0.013$
to $+0.022$; the seed whose realized $m$ fell by $1.1$ changed by
$-0.069$.
All twelve seeds with $\Delta\bar m<-0.3$ changed negatively, while
$60\%$ of the seeds changed positively.
At C, the same nominal operating point therefore produced feedback
responses of both signs across seeds.

\section{Mechanisms and robustness}
\label{sec:Mech}

\subsection{Microscopic reorganization across the crossover}
\begin{figure}[!htbp]
  \centering
          \includegraphics[width=1\textwidth]{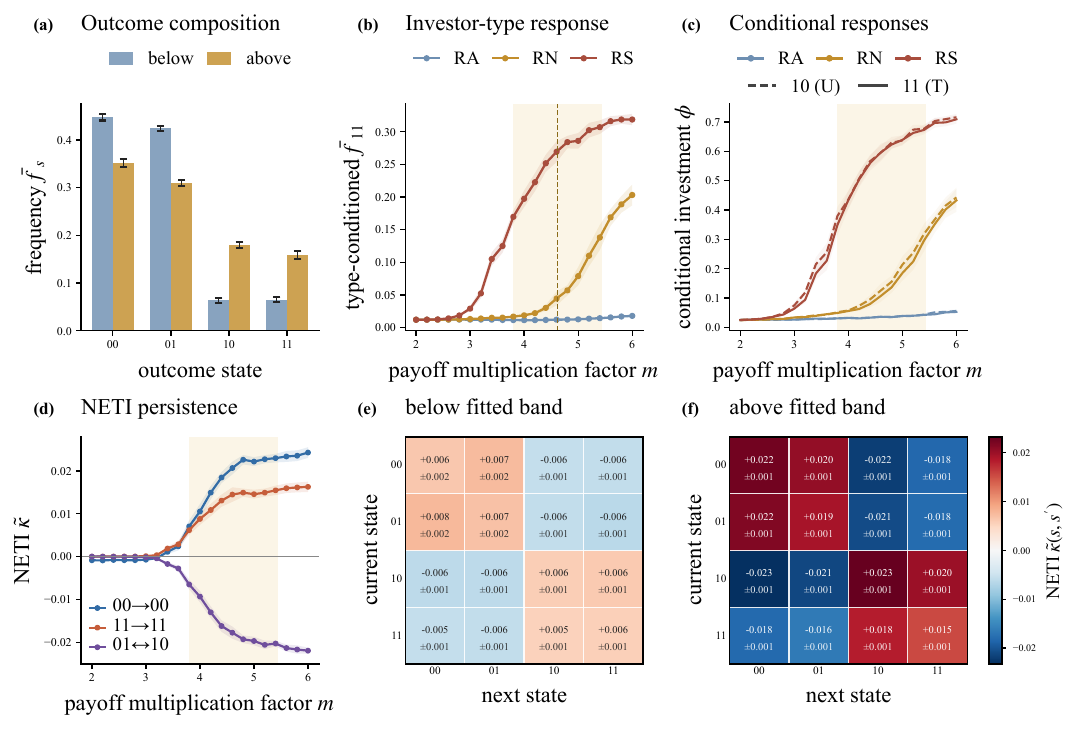}
  \caption{Finite-horizon mechanism atlas at $\rho=0.2075$. (a) Outcome-state
  composition below ($m^*-w=3.789$) and above ($m^*+w=5.444$) the nearest
  fitted crossover. (b) Type-conditioned successful trust across $m$.
  (c) Type-resolved investment conditional on the previous outcome:
  dashed lines denote outcome 10 (untrustworthy, $U$), and solid lines
  denote outcome 11 (trustworthy, $T$). In (b,c), RA, RN, and RS denote
  risk-averse, risk-neutral, and risk-seeking investors.
  (d) Three NETI signatures across $m$: $00\!\to\!00$,
  $11\!\to\!11$, and the mean $01\!\leftrightarrow\!10$ cross-block entry.
  (e,f) Full NETI matrices below and above the fitted band, with one shared
  color scale and cellwise confidence intervals. Curves, bars, and ribbons
  are means and two-sided 95\% Student-$t$ confidence intervals across 20
  independent static-baseline seeds over rounds 3,000--11,999 of the fixed
  $T=12{,}000$ horizon. The gold region denotes the fitted $m^*\!\pm w$
  band.}
  \label{fig:mechanism-atlas}
\end{figure}

We examined investor participation and outcome composition across the fitted
crossover. Across this band, the non-investment outcomes
00 and 01 decreased from $0.447\pm0.006$ and $0.424\pm0.005$ to
$0.351\pm0.009$ and $0.309\pm0.006$, whereas 10 and 11 increased from
$0.064\pm0.005$ and $0.065\pm0.005$ to $0.180\pm0.006$ and $0.159\pm0.008$
(Fig.~\ref{fig:mechanism-atlas}a). The type-resolved change was concentrated
in RN and RS investors. RN $\bar f_{11}$ increased from $0.017\pm0.001$ to
$0.148\pm0.015$ and RS $\bar f_{11}$ from $0.165\pm0.012$ to
$0.312\pm0.008$; RA remained low ($0.011\pm0.000$ to $0.015\pm0.001$).
Conditional investment frequencies after outcomes 10 and 11, respectively,
increased from $(0.050,0.048)$ to $(0.333,0.306)$ for RN and from
$(0.366,0.328)$ to $(0.689,0.677)$ for RS, while the corresponding RA values
changed only from $(0.031,0.031)$ to $(0.046,0.044)$
(Fig.~\ref{fig:mechanism-atlas}b,c). Values are means $\pm$ 95\% Student-$t$
confidence interval half-widths; the smallest per-seed conditional
denominator was 6,247. RS investors lead the response, with an increasing
RN contribution at larger $m$.

\subsection{Pooled outcome-pair structure: the NETI view}

We compared NETI across values of the payoff multiplication factor $m$
and found stronger excess self-pairing for both 00 and 11 across the crossover.
At $m=2.0$, $\tilde\kappa(00,00)=-0.0009\pm0.0001$ and
$\tilde\kappa(11,11)$ was approximately zero. By $m=4.6$, the two entries
had increased to $0.021\pm0.001$ and $0.014\pm0.002$, respectively, while
the mean $01\!\leftrightarrow\!10$ entry declined to $-0.018\pm0.001$. At
$m=6.0$, the corresponding values were $0.024\pm0.001$, $0.016\pm0.001$, and
$-0.022\pm0.001$. The matrices below and above the fitted crossover band
showed the same pattern:
$\tilde\kappa(00,00)$ increased from $0.006\pm0.002$ to $0.022\pm0.001$, and
$\tilde\kappa(11,11)$ from $0.006\pm0.001$ to $0.015\pm0.001$
(Fig.~\ref{fig:mechanism-atlas}d--f). These NETI changes describe pooled
population-level pair frequencies. Values are means
$\pm$ 95\% Student-$t$ confidence interval half-widths over 20 seeds.

\subsection{Myopic thresholds and type-resolved participation}
\label{sec:Thresholds}

We define $\hat q$ as the perceived trustworthiness probability and derive
a myopic benchmark from immediate subjective payoffs.
For type $\theta$, investing has subjective value
$\hat q\,(m/2-1)-(1-\hat q)\lambda_\theta$ for $m\ge 2$, giving the risk-adjusted trust threshold
\begin{equation}
  \tilde q^{\,*}_\theta(m)=\frac{2\lambda_\theta}{m+2\lambda_\theta-2},
  \qquad
  \tilde q^{\,*}_{\mathrm{RA}}>\tilde q^{\,*}_{\mathrm{RN}}>\tilde q^{\,*}_{\mathrm{RS}}
  \quad (m>2).
  \label{eq:trust-threshold-theory}
\end{equation}
At $m=2$ every type shares the threshold $\tilde q^{\,*}_\theta=1$.
We use this static benchmark to predict the participation ordering
(\ref{app:threshold-derivation}).

For a trustee receiving positive inflow $I_j>0$, the untrustworthy action
yields an immediate monetary advantage $\rho m I_j/2$.
The trustworthy action instead yields an additional reputation increment
$\eta_R$, independent of the current reputation (\ref{app:threshold-derivation}).
Trustworthiness is therefore weakly preferred in the myopic comparison when
\begin{equation}
  \lambda_\phi \;\ge\; \lambda_\phi^{*}=\frac{\rho\,m\,I_j}{2\eta_R}.
  \label{eq:trustworthy-threshold}
\end{equation}
At zero inflow, the two actions are myopically indifferent.
On the fixed-$\rho_0=0.2075$ slice, the crossing
$m_{\rm cross}=2\eta_R\lambda_\phi/(\rho_0 I_j)$ lies below $2$ for both
trustee types in Table~\ref{tab:parameters}, even at $I_j=1$.
This predicts no trustee-type onset within $m\in[2,6]$ on that slice.
Targeted checks with $\eta_R=0.2$ or $\lambda_L=5$ also failed to produce
a theory-aligned L--S trustworthiness response.

At $m=2$, all three investor types remained near a common
exploration-dominated floor ($\bar f_{11}\simeq0.012$). By $m=4.6$,
type-conditioned $\bar f_{11}$ was $0.012\pm0.001$, $0.044\pm0.011$, and
$0.269\pm0.016$ for RA, RN, and RS, respectively (mean $\pm$ 95\%
confidence-interval half-width). At the same $m$, investment after an
untrustworthy outcome (10) was $0.036\pm0.002$, $0.126\pm0.023$, and
$0.595\pm0.025$, and investment after a successful outcome (11) was $0.036\pm0.002$, $0.107\pm0.020$, and $0.596\pm0.030$.
RS led the rise through most of the sampled range, RN followed at larger
$m$, and RA remained low. This ordering emerged above the shared exploration
floor (Fig.~\ref{fig:mechanism-atlas}b,c).

\subsection{Continuation value advances the risk-seeking onset}
\label{sec:OnsetBenchmark}

\begin{figure}[!htbp]
  \centering
            \includegraphics[width=1\textwidth]{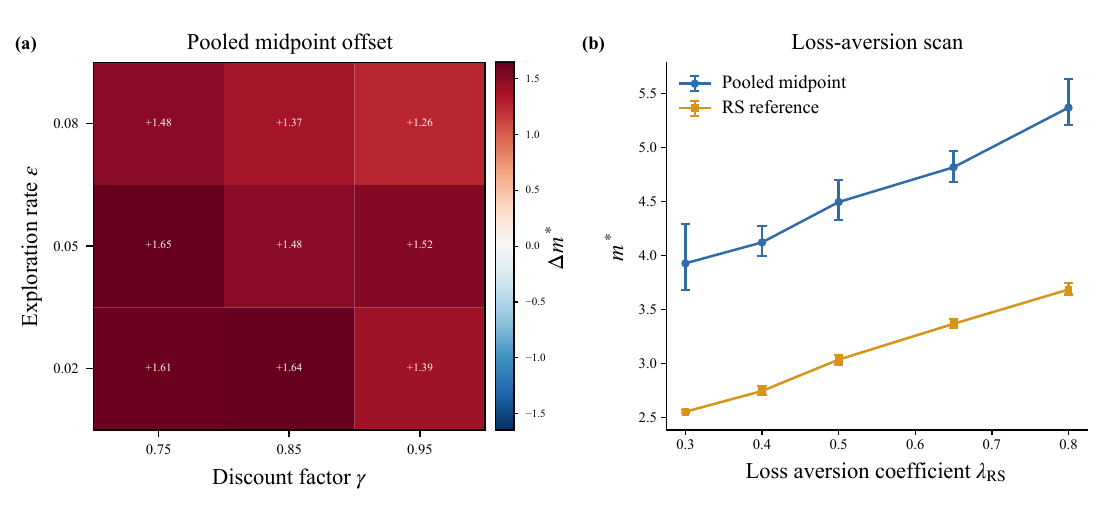}
  \caption{Pooled fitted midpoints and the RS one-round reference. (a)
  Offset $\Delta m^*=m_{\mathrm{fit}}^*-m_{\mathrm{theory}}^*$ across the
  $(\gamma,\varepsilon)$ plane. (b) Pooled midpoints and RS references across the
  different loss-aversion coefficients $\lambda_{\mathrm{RS}}$. Error bars are two-sided 95\%
  whole-seed bootstrap confidence intervals from 1,000 resamples. For every
  setting, $q_0=N_{11}/(N_{10}+N_{11})$ is measured independently at fixed
  $m=3.0$ from 20 seeds over rounds 3,000--11,999; all panels use the static
  baseline at $\rho=0.2075$. The fitted midpoint summarizes the three-type
  population mixture, whereas the reference is the RS one-round utility zero.
  Both increase with the RS loss aversion coefficient.}
  \label{fig:theory-validation}
\end{figure}

We measured conditional trustworthiness $q_0$ in the low-$m$ region, away from
the fitted crossover. Inverting Eq.~\eqref{eq:trust-threshold-theory} at this
value gives the static onset benchmark
\begin{equation}
  m_1^{\,*}(\lambda,q_0)=2+\frac{2\lambda(1-q_0)}{q_0},
  \label{eq:m1-theory}
\end{equation}
which we evaluate for the risk-seeking type at
$m_1^{\,*}(\lambda_{\mathrm{RS}},q_0)$. Across the nine
$(\gamma,\varepsilon)$ settings, the independently measured
$q_0=0.486$--$0.502$ gave benchmark onsets
$m_{\mathrm{theory}}^*=2.99$--$3.06$, whereas logistic fits to the pooled
three-type response gave midpoints $m_{\mathrm{fit}}^*=4.25$--$4.70$. The offset
$m_{\mathrm{fit}}^*-m_{\mathrm{theory}}^*$ was therefore positive in every
cell ($+1.26$ to $+1.65$), with all 95\% whole-seed bootstrap intervals
excluding zero (Fig.~\ref{fig:theory-validation}a).

Increasing $\lambda_{\mathrm{RS}}$ from 0.30 to 0.80 raised both the pooled
fitted midpoint ($3.93\to5.37$) and the RS benchmark ($2.56\to3.69$). Their offset
remained $+1.37$ to $+1.68$, with all bootstrap intervals above zero
(Fig.~\ref{fig:theory-validation}b). The pooled midpoint and the RS one-round
reference thus shifted in the same direction as loss aversion increased.

We tested the contribution of investor continuation value by setting
$\gamma_{\mathrm{inv}}=0$ while retaining $Q$-table updates
(\ref{app:ident-myopic}, Fig.~\ref{fig:ii5-decomp}). We compared both RS fitted
midpoints with a common one-round reference estimated at
$\gamma_{\mathrm{inv}}=0.9$. The zero-discount midpoint was
$4.24\pm0.08$, a $+1.21\pm0.08$ gap above this reference,
$m_1^{\,*}=3.035\pm0.031$. Restoring continuation value lowered the
midpoint to $3.61\pm0.04$, a paired shift of $-0.63\pm0.08$
(estimates $\pm$ SE). Continuation value therefore advanced the RS crossover.

\subsection{A quasi-static description of the payoff response}
\label{sec:LoopGain}

Centering sets the environmental drive to zero at the reference observables.
We construct a quasi-static scalar approximation from smooth static-response
maps within unchanged payoff and observation tiers (\ref{app:feedback-derivation}).
Its dimensionless loop gain is
\begin{equation}
  G_{\mathcal W}(m_0,\rho_0)=M'(E_0)\left[\beta_C\chi^{\mathcal W}_{C,m}
  +\beta_r\chi^{\mathcal W}_{r,m}\right],
  \label{eq:loop-gain-main}
\end{equation}
where $M'(E_0)=k(m_{\max}-m_{\min})z_0(1-z_0)$ is the sigmoid slope.
The susceptibilities $\chi^{\mathcal W}_{C,m}$ and $\chi^{\mathcal W}_{r,m}$
are derivatives of the smooth approximations to finite-window successful
trust and mean investor payoff.
For a stable scalar map, a sustained forcing is amplified by
$1/(1-G_{\mathcal W})$.
Its discrete multiplier is $a=1-\nu(1-G_{\mathcal W})$; for $0<a<1$,
the relaxation time is $\tau=-1/\log a\simeq1/[\nu(1-G_{\mathcal W})]$.

We applied a small environmental impulse in a checkpoint-fork control
(\ref{app:ident-amplifier}, Fig.~\ref{fig:ii3-amplifier}). At the mid-band
operating point, the response decayed with a fitted $\tau=41.45$ rounds (SE 3.51).
The discrete-time conversion gave $G_{\mathrm{eff}}=0.205$
(approximate 95\% CI $[0.020,0.332]$). In the scalar model,
this positive gain slows the decay of environmental perturbations.

The $m$-only change in successful trust agreed in sign with the static-baseline
benchmark at 10 of 11 selected nominal points (\ref{app:feedback-derivation}).
We obtained the benchmark change by multiplying the baseline finite-difference
slope by the realized mean excursion $\Delta\bar m$.
The median observed-to-benchmark ratio was $0.42$
(interquartile range $[0.252,0.642]$), so the observed changes were
typically smaller in magnitude than the finite-grid benchmark.

\subsection{Perception responses near tier boundaries}
\label{sec:PerceptionMechanism}

The largest positive slice responses occurred near $m_0=3.3$ and $4.7$, each only
$0.033$ from a tier edge, compared with temporal standard deviations of
$0.05$--$0.12$ in realized $m(t)$ (Table~\ref{tab:slice-channels}).
At $m_0=3.3$, the $\rho$-only branch changed perception while both payoff
parameters stayed fixed. Its positive response was comparable in size to the
$m$-only response. At $4.7$, both payoff parameters also stayed nominal,
but the positive point estimate had an interval spanning zero. At $4.5$,
the nearest edge was five times farther away and the mean excursion was
directed away from it; both branch intervals spanned zero.

First entry into a tier absent from the fixed-environment burn-in activates
unupdated $Q$ rows. Uniform tie-breaking gives each binary action probability
$1/2$ on its first selection, which can initially raise investment above a
low learned rate. The full-window contrast combines first visits, revisits,
and subsequent learning.

We separated payoff and perception effects in a $2\times2$ factorial
experiment (\ref{app:ident-channels}, Fig.~\ref{fig:ii2-channels}).
Perception responses were nonzero in individual runs near tier boundaries,
whereas dynamic and fixed perception produced identical successful-trust
trajectories at the tested mid-band point. The interaction interval spanned
zero and effects of either sign comparable in magnitude to the main effects.
At $m_0=3.333$, the perception response was nonzero under
$\kappa=(1/3,2/3)$ and exactly zero under $\kappa=(0.25,0.75)$
(\ref{app:ident-edges}, Fig.~\ref{fig:ii6-edges}).
These results identify the observed environmental tier as a route to
behavioral change. The between-scheme comparison combines changes in state
discretization, the $\rho$-switching rule and the resulting learning histories.

\subsection{Feedback preserves the pooled two-block sign pattern at B--D}
\begin{figure}[!htbp]
  \centering
              \includegraphics[width=1\textwidth]{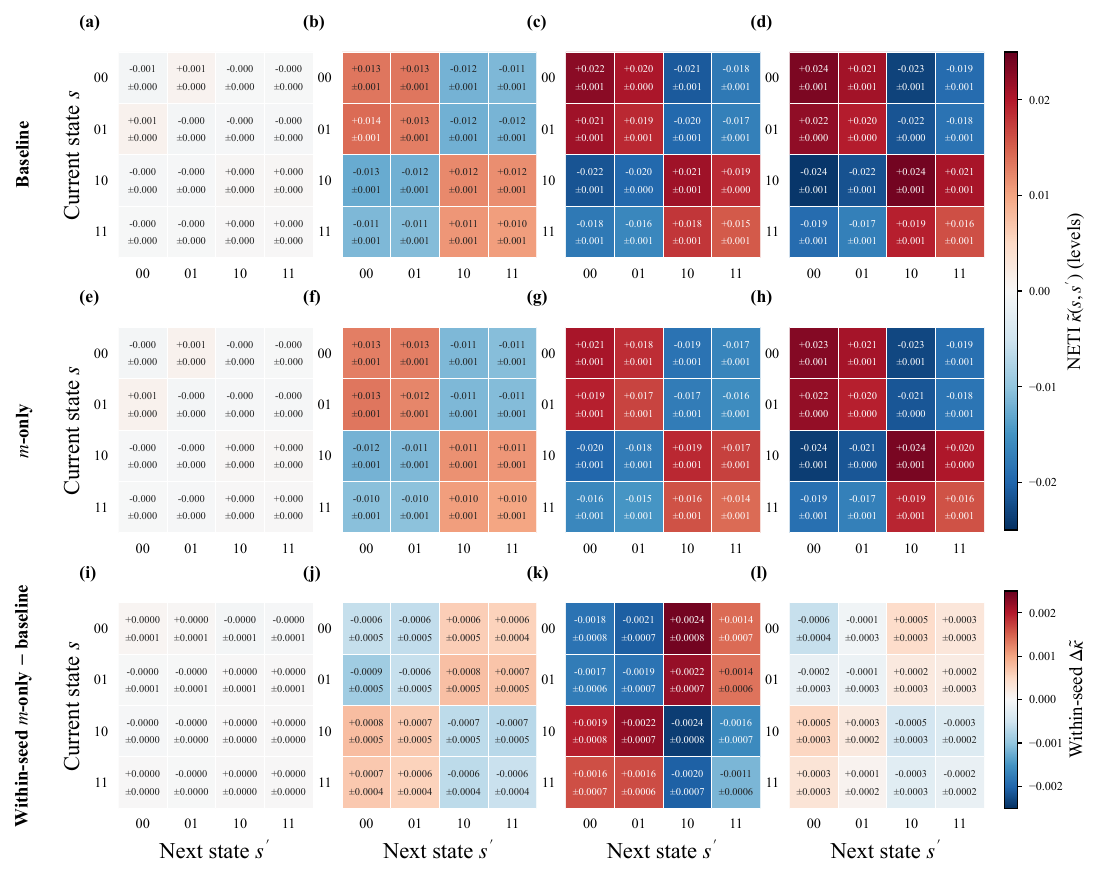}
  \caption{Finite-window NETI matrix comparison at the four parameter settings.
  Panels (a--d), (e--h), and (i--l) form the baseline, $m$-only, and
  within-seed $m$-only-minus-baseline rows, respectively; columns follow
  A--D, with $m_0=3.1$, 4.1, 4.9, and 5.7, respectively, at
  $\rho_0=0.2075$. The first two rows share a symmetric level scale and the contrast row
  has its own symmetric difference scale, ten times finer. Cell text reports
  the mean and two-sided 95\% Student-$t$ confidence interval half-width
  across 50 matched seeds over $[3000,12000)$. NETI reports finite-window,
  cellwise joint-frequency excess relative to a shuffled benchmark.}
  \label{fig:selected4-neti-matrix-grid}
\end{figure}

Under feedback, the pooled NETI matrices retained their principal two-block
sign pattern at B--D, while A remained weakly structured near the low-trust
level (Fig.~\ref{fig:selected4-neti-matrix-grid}). Across the 64 paired NETI
cells, mean differences ranged from $-0.0024$ to $+0.0024$; 21
pointwise intervals lay above zero, 20 below zero, and 23 included zero.
The largest paired difference at each point was $4\%$, $6\%$, $11\%$, and
$2\%$ of the largest baseline entry at A--D.

NETI retained the two-block sign pattern at B--D under conditional
independence benchmarks formed from population outcome margins within
100-round blocks or each adjacent pair of rounds (\ref{app:kappa-details}).
A remained weakly structured near its low-trust level.

\subsection{Robustness of the finite-horizon feedback response}
\label{sec:Robust}
\begin{figure}[!htb]
  \centering
                \includegraphics[width=1\textwidth]{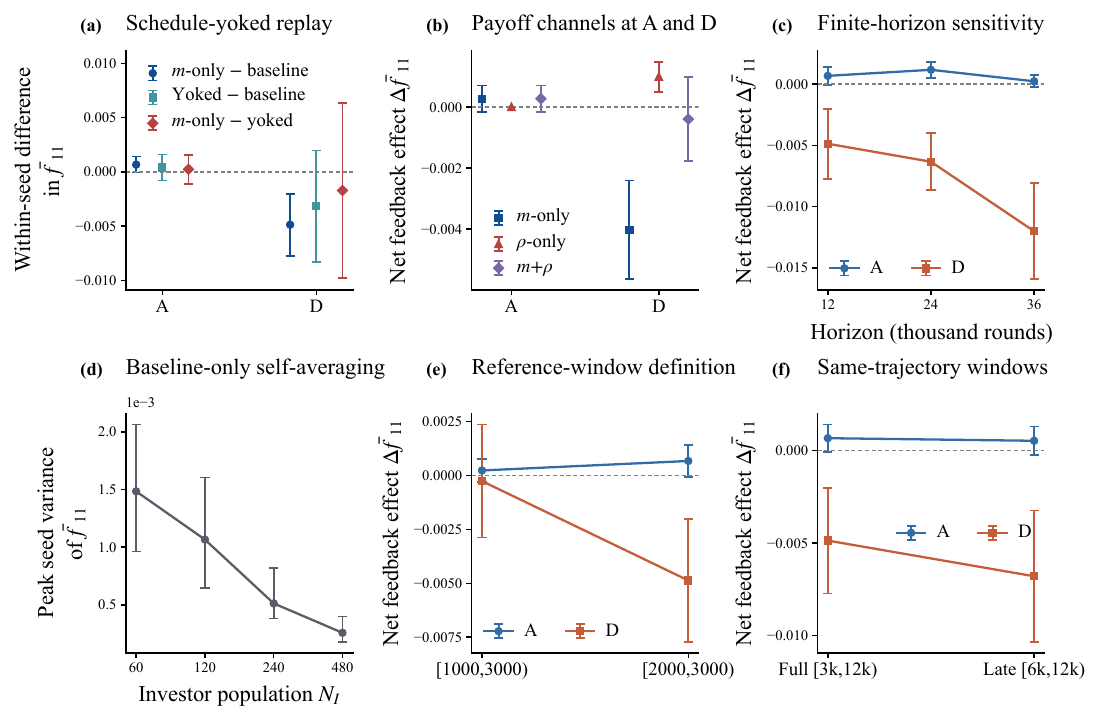}
  \caption{Controls on the centered finite-horizon response at points A and D.
  (a) Within-seed $m$-only-minus-baseline, foreign-schedule-yoked-minus-baseline,
  and $m$-only-minus-yoked contrasts. (b) Within-seed changes of
  the $m$-only, $\rho$-only, and joint-channel branches relative to the
  baseline; the $\rho$-only change at D is positive while the $m$-only
  change is negative. (c) Paired
  $m$-only changes at 12k, 24k, and 36k rounds, each summarized over
  $[3000,T)$. (d) Baseline-only peak seed variance across four population
  sizes. (e) Net feedback effects under reference windows $[1000,3000)$ and
  $[2000,3000)$. (f) Full $[3000,12000)$ and fixed late $[6000,12000)$
  observation windows using the reference window $[2000,3000)$. Panels (a,c,e,f) use 20
matched seeds. Intervals involving yoked replay in (a) use the alternating-group
  Bonferroni procedure in \ref{app:parameters} (approximate pointwise 95\% coverage).
  The other intervals in (a,c,e,f) are $t_{19}$ intervals; panel (b) uses
  50 matched seeds and $t_{49}$ intervals. Panel (d) uses 20 baseline
  seeds and 1,000 whole-seed bootstrap replicates. All intervals are pointwise.}
  \label{fig:feedback-controls}
\end{figure}

\begin{figure}[!htb]
  \centering
   \includegraphics[width=1\textwidth]{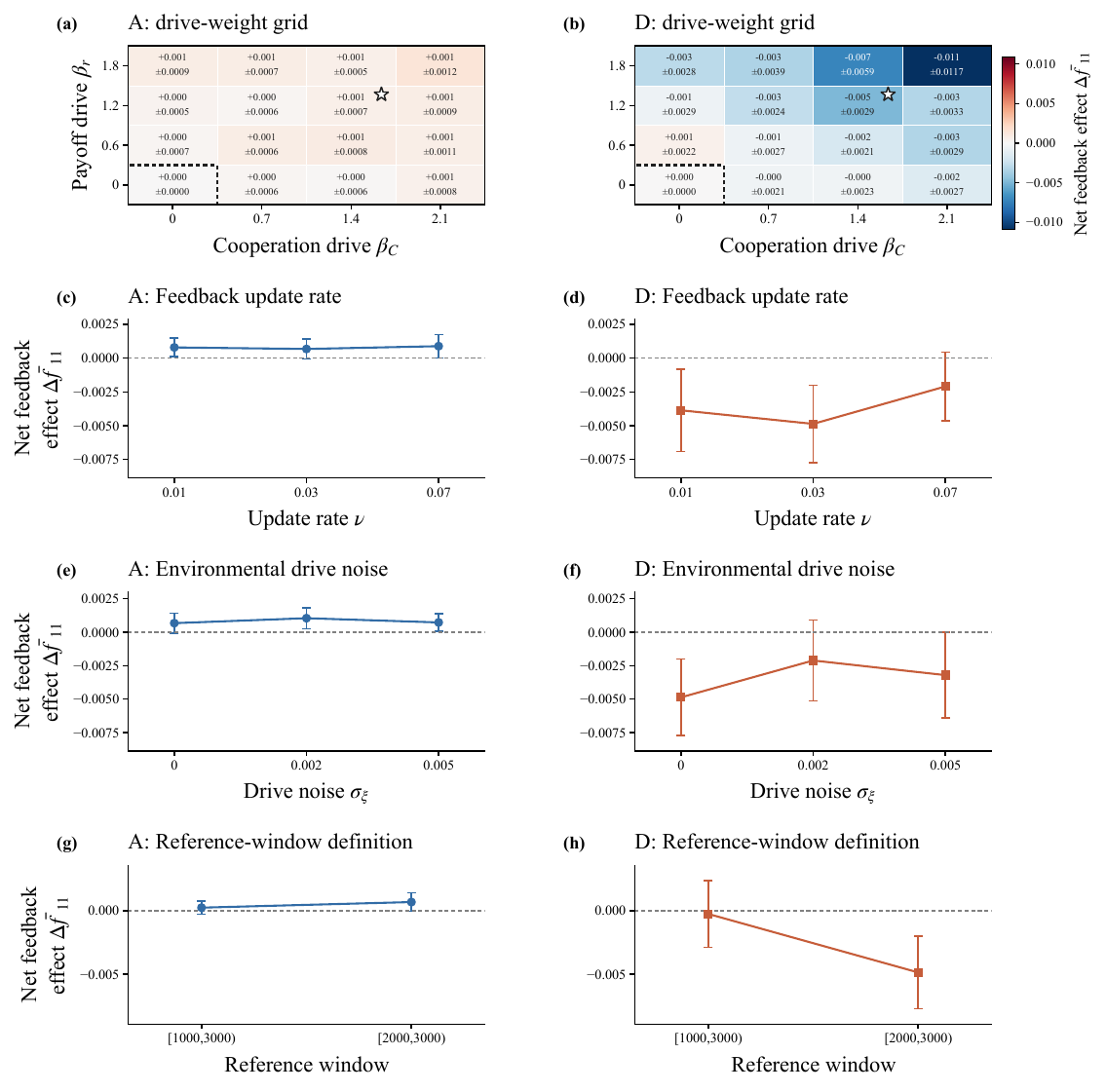}
  \caption{Paired $m$-only sensitivity at A (left) and D (right).
  (a,b) Responses over the $4\times4$ drive-weight grid; dashed outlines mark
  the zero-drive cells ($\beta_C=\beta_r=0$), and stars mark
  $(\beta_C,\beta_r)=(1.4,1.2)$.
  (c,d) Environmental update rate. (e,f) Drive noise.
  (g,h) Reference-window definition. Entries and error bars show means and
  pointwise 95\% Student-$t$ intervals across 20 matched seeds over
  $[3000,12000)$.}
  \label{fig:feedback-sensitivity-grid}
\end{figure}

With $\beta_C=\beta_r=0$, we obtained the same successful-trust
trajectory as in the static baseline (Figs.~\ref{fig:feedback-controls}
and~\ref{fig:feedback-sensitivity-grid}).
Along the diagonal of increasing
drive weights $(\beta_C,\beta_r)=(0.7,0.6)$, $(1.4,1.2)$, $(2.1,1.8)$, the
net feedback effect at D was $-0.0006$, $-0.0049$, and $-0.011$; at A, on the
floor, it was $+0.0008$, $+0.0007$, and $+0.0014$
(Fig.~\ref{fig:feedback-sensitivity-grid}a,b). Across the 16 cells, five
pointwise intervals at D lay below zero and five at A above zero; the
interval at the main drive weights included zero at A and lay below zero at D.

At D, extending the horizon made the net feedback effect more negative,
from $-0.0049$ at 12k to
$-0.0063$ at 24k and $-0.012$ at 36k rounds. All three intervals were below
zero. Restricting the observation window to $[6000,12000)$ changed the effect
from $-0.0049$ to $-0.0068$ (Fig.~\ref{fig:feedback-controls}c,f). At A the 12k
and 36k intervals included zero and the 24k mean was $+0.0012$. The payoff
response thus continued to accumulate over the tested horizons at D, where
the baseline learning decline also persisted.

Extending the reference window from $[2000,3000)$ to
$[1000,3000)$ includes the rising part of the trajectory and therefore lowers
the reference average. With this longer reference window, the net feedback effect at D changed from $-0.0049$
(95\% CI $-0.0077$ to $-0.0020$) to $-0.0003$ ($-0.0029$ to $+0.0024$;
Fig.~\ref{fig:feedback-sensitivity-grid}h), attenuating the estimated response
toward zero.

Varying the environmental update rate $\nu$ over $0.01$, $0.03$, and $0.07$
gave net feedback effects of $-0.0039$, $-0.0049$, and $-0.0021$ at D,
respectively (Fig.~\ref{fig:feedback-sensitivity-grid}d).
Adding drive noise $\sigma_\xi=0.002$ or $0.005$ gave $-0.0021$ and $-0.0032$
at D, respectively (Fig.~\ref{fig:feedback-sensitivity-grid}f).
At A, the effects remained between $+0.0007$ and $+0.0010$ under both variations
(Fig.~\ref{fig:feedback-sensitivity-grid}c,e). The estimated net feedback effects at D remained negative across these
settings, with some confidence intervals including zero.

At D the mean $m$-only, $\rho$-only, and joint changes relative to the
baseline were $-0.0040$, $+0.0010$, and $-0.0004$
(Fig.~\ref{fig:feedback-controls}b). The $\rho$-only point estimate has the
opposite sign to the $m$-only estimate, and the joint estimate lies between
them. For these branches, over $[3000,12000)$ and the 50 matched
seeds, the paired departure from additivity
$\Delta\bar f_{11}^{m+\rho}-(\Delta\bar f_{11}^{m}+\Delta\bar f_{11}^{\rho})$
is $+0.0026$ (95\% CI $[+0.0018,+0.0034]$). Each branch evolves its own
environment and perceived state; the factorial control in
\ref{app:ident-channels} instead switches payoff and perception separately.
The combined branch response at D differed from the sum of the two
separate responses over this window.

We used schedule-yoked replay to apply each seed's environmental schedule
to the preceding seed's learners, forming a cycle of 20 recipient--donor pairs.
Relative to baseline, the $m$-only contrasts were
$+0.0007$ at A (95\% CI $[-0.0001,+0.0014]$)
and $-0.0049$ at D ($[-0.0077,-0.0020]$).
The corresponding yoked contrasts were $+0.0004$
($[-0.0008,+0.0016]$) and $-0.0032$
($[-0.0083,+0.0020]$; approximate 95\% CIs).
The direct $m$-only-minus-yoked contrast was $+0.0003$
at A (approximate 95\% CI $[-0.0011,+0.0016]$) and
$-0.0017$ at D ($[-0.0098,+0.0063]$).
The intervals account conservatively for shared donors using alternating
groups (\ref{app:parameters}). Both direct-comparison intervals included zero
(Fig.~\ref{fig:feedback-controls}a).

Across $N_I=60,120,240,480$, the fitted crossover width of the static
baseline decreased from $0.96$ to $0.73$ (95\% whole-seed bootstrap CIs
$[0.84,1.07]$ and $[0.68,0.79]$), and the variance peak decreased from
$0.0015$ to $0.00026$ ($[0.00096,0.0021]$ and $[0.00018,0.00040]$;
Fig.~\ref{fig:feedback-controls}d). The shared-seed bootstrap slopes against
$\log_2N_I$ were $-0.072$ $[-0.11,-0.033]$ for the width and $-0.00041$
$[-0.00060,-0.00023]$ for the variance peak. Both trends are resolved over
the sampled sizes at the fixed 12,000-round horizon: larger populations
showed a narrower crossover and lower between-seed variability.

\section{Discussion and conclusion}
\label{sec:Con}
We find an asymmetric trust response to environmental feedback: trust increases
near perception-tier boundaries, whereas feedback reinforces learning-driven
trust erosion at the high-return operating point D.
At $m_0=3.3$, changing the perceived environmental tier raised successful trust
from 3.2\% to 4.3\% while both payoff parameters remained fixed.
At D, stronger drive and longer horizons deepened the negative response,
whereas extending the historical reference window backward attenuated it.
The direction of feedback therefore depends on the learners' states and
histories; stronger feedback does not consistently produce higher trust.

We couple a centered environmental update with heterogeneous $Q$-learning
to close the loop between collective behavior, environmental change, and
subsequent decisions. Relative to the fixed-payoff trust game, collective
outcomes now reshape both material incentives and publicly perceived conditions.
Paired branches from a shared learning checkpoint and separate payoff and
perception controls identify the trust responses through these two routes.
Previous work reported high trust in repeated $Q$-learning games with
alternating roles~\cite{Zheng2024DecodingTrust}; our model examines how collective
learning changes its own decision environment under fixed roles.

Behavioral heterogeneity links this collective response to individual learning.
Risk-seeking investors lead participation, the one-round thresholds explain
the RS--RN--RA ordering, and the discount-factor control shows that continuation
value advances the RS onset.
At the population level, feedback changes successful trust while preserving
the pooled two-block NETI sign pattern at B--D.
Trust levels can thus change within a retained organization of outcome pairs.

In our tabular learner, changes in the perceived tier activate different $Q$ rows,
and the response combines first visits to untrained rows with subsequent learning.
This links the perception effect to state representation and learning history.
These findings offer an analogy for platforms at different stages of development.
In a community's cold-start phase, perceived public conditions could provide
an entry point for building trust.
In a mature platform, high returns alone may be insufficient to sustain trust
as learning accumulates.
This analogy motivates interventions differentiated by trust state and proximity
to perception-tier boundaries.
For low-trust communities near a boundary, one testable intervention is public
feedback that makes improvements in collective cooperation visible.
For high-return platforms with declining trust, a complementary test is whether
constraining returns to defection or introducing penalties can curb trust erosion.
These are hypotheses motivated by the asymmetric model response, to be
tested under the information and learning conditions of the platform.

Our results concern two populations with fixed roles, random matching each
round, and finite observation horizons.
Extending the model to dynamic networks with partner choice and rewiring would
establish how evolving relationships alter the payoff and perception responses.
Within the present model, trust maintenance depends jointly on material
incentives, public state representation, and collective learning history.

\section*{CRediT authorship contribution statement}
\textbf{Ruqiang Guo:} Conceptualization, Methodology, Writing -- original draft.\par
\textbf{Linjie Liu:} Conceptualization, Writing -- review \& editing.\par
\textbf{Fangfang Wang:} Investigation, Formal analysis.\par
\textbf{Zhaoyi Hu:} Formal analysis, Discussing, \& Investigation.\par

\section*{Declaration of Competing Interest}
The authors declare that they have no known competing financial interests or personal relationships that could have
appeared to influence the work reported in this paper.

\section*{Acknowledgment}
We would like to thank the referees for their careful reading and helpful comments. L. Liu’s research was supported by the Natural Science Foundation of Shaanxi (Grant No. JC-QN-0791) and the Fundamental Research Funds of the Central Universities of China (Grant Nos. 2452022012, 2452022144). R. Guo’s research was supported in part by the Project for Enhancing Young and Middle-aged Teachers' Basic Scientific Research Ability in Universities of Guangxi (Grant No. 2025KY1899).


\clearpage
\appendix

\section{Theoretical analysis and derivations}
\label{app:mech-details}

\subsection{NETI properties}
\label{app:kappa-details}
Let $X_i(t)\in\mathcal S_o:=\{00,01,10,11\}$ be investor $i$'s realized
outcome state in round $t$. For a contiguous analysis window $\mathcal W$, we
count adjacent pairs
\begin{equation}
 N_{ss'}=\sum_{i=1}^{N_I}\sum_{t\in\mathcal W}
 \mathbf 1\!\left[X_i(t-1)=s,\,X_i(t)=s'\right],\qquad
 P_o(s,s')=\frac{N_{ss'}}{N_I|\mathcal W|}.
 \label{eq:neti-joint}
\end{equation}
We also compute the pooled current-state frequency
$\pi_+(s)=\sum_{s'}P_o(s',s)$, which equals the outcome share used in
Eq.~\eqref{eq:finite-window-mean}, and use the symmetric shuffled benchmark
$P_e(s,s')=\pi_+(s)\pi_+(s')$ in Eq.~\eqref{eq:kappa}. The previous-state
marginal $\pi_-(s)=\sum_{s'}P_o(s,s')$ can differ from $\pi_+(s)$ because the
window replaces its initial boundary state by its final boundary state. For
each $s$ this discrepancy is bounded by $1/|\mathcal W|$.

For comparison, the empirical conditional transition probability is defined
when $\pi_-(s)>0$ as
\begin{equation}
 K_{\mathcal W}(s'\mid s)=\frac{P_o(s,s')}{\pi_-(s)},
 \label{eq:window-transition-probability}
\end{equation}
whereas NETI measures normalized excess joint frequency relative to this shuffled
benchmark, retaining the exposure frequency of state $s$.
A positive cell means $P_o>P_e$, a negative cell means
$P_o<P_e$, and a zero cell means equality with this finite-window benchmark.

The attainable upper bound depends on the cell marginals. Every joint table obeys
\[
P_o(s,s')\leq\min\{\pi_-(s),\pi_+(s')\}.
\]
Consequently,
\begin{equation}
 \tilde\kappa(s,s')\leq
 \frac{\min\{\pi_-(s),\pi_+(s')\}-\pi_+(s)\pi_+(s')}
 {1-\pi_+(s)\pi_+(s')},
 \label{eq:neti-upper-bound}
\end{equation}
whenever the denominator is positive. In the limiting degenerate case
$P_e=1$, the formula is undefined and we assign zero by convention.
Under equal row and column
marginals, Eq.~\eqref{eq:neti-upper-bound} is strictly below one for every
nondegenerate off-diagonal cell (and equals $\pi(s)/(1+\pi(s))$ on the
diagonal). We therefore compare NETI magnitudes across matched conditions,
using seed-level uncertainty and the cell-specific bounds.

We also formed conditional independence benchmarks from the population
outcome margins within non-overlapping blocks of $L$ rounds.
Let $n_t(s)=\sum_{i=1}^{N_I}\mathbf1[X_i(t)=s]$ and let $\mathcal B_L$
partition $\mathcal W=[3000,12000)$ into equal blocks. The analytic
independent-pairing expectation is
\[
 P_e^{(L)}(s,s')=
 \frac{\displaystyle\sum_{B\in\mathcal B_L}
  \left[\sum_{t\in B}n_{t-1}(s)\right]
  \left[\sum_{t\in B}n_t(s')\right]}
 {N_I^2 L|\mathcal W|}.
\]
We used $L=100$ to preserve each block's lagged and current outcome margins,
and $L=1$ to preserve the exact margins of each adjacent-round pair.
For each seed and branch, we substituted $P_e^{(L)}$ for $P_e$ in
Eq.~\eqref{eq:kappa}, retaining the observed full-window $P_o$.

\subsection{Threshold derivations}
\label{app:threshold-derivation}
For an investor who commits one unit, a trustworthy trustee produces the
monetary payoff $r^{T}=m/2-1$ and an untrustworthy one produces
$r^{U}=-1$. For
$m\geq2$, the former lies on the non-negative branch of
Eq.~\eqref{eq:utility-investor}. In this myopic benchmark, we denote perceived
trustworthiness by $\hat q$. The one-round expected subjective utility of investing
is
\begin{equation}
 U_\theta(1\mid\hat q)=\hat q\left(\frac m2-1\right)
 -(1-\hat q)\lambda_\theta ,
 \end{equation}
whereas not investing yields zero. Solving
$U_\theta(1\mid\hat q)\geq0$ gives
\begin{equation}
 \hat q\geq\tilde q_\theta^{\,*}(m)
 =\frac{\lambda_\theta}{m/2-1+\lambda_\theta}
 =\frac{2\lambda_\theta}{m+2\lambda_\theta-2}.
 \label{eq:investor-threshold-derivation}
\end{equation}
For $m>2$, this threshold decreases with $m$ and increases with
$\lambda_\theta$, yielding the ordering in
Eq.~\eqref{eq:trust-threshold-theory}; at $m=2$ it equals one for every
$\lambda_\theta>0$.
The simulated agents choose actions from learned $Q$-values with constant
$\varepsilon$-greedy exploration.

For a trustee with positive inflow $I_j>0$, let $R_j$ be its reputation before
the action. The two possible next reputations are
\begin{align}
 R_j^{T}(t+1)&=(1-\eta_R)R_j(t)+\eta_R,\\
 R_j^{U}(t+1)&=(1-\eta_R)R_j(t).
\end{align}
Although the action-specific changes are
$\eta_R(1-R_j)$ and $-\eta_RR_j$, respectively, their difference is exactly
\begin{equation}
 \Delta R_j^{T}-\Delta R_j^{U}=\eta_R,
 \label{eq:reputation-differential}
\end{equation}
independent of the current reputation. Meanwhile the immediate monetary
advantage of the untrustworthy action is
\begin{equation}
 w_j^{U}-w_j^{T}=\frac{\rho m I_j}{2}.
\end{equation}
The one-round utility difference between the trustworthy and the
untrustworthy action is hence $\lambda_\phi\eta_R-\rho m I_j/2$, and the
trustworthy action is weakly preferred under the myopic comparison when
\begin{equation}
 \lambda_\phi\geq\lambda_\phi^*(m,\rho,I_j)
 =\frac{\rho m I_j}{2\eta_R}.
 \label{eq:trustee-threshold-derivation}
\end{equation}
When $I_j=0$, both actions have the same monetary payoff and both set the
reputation signal to zero, so they are myopically indifferent; the differential
in Eq.~\eqref{eq:reputation-differential} applies only to positive inflow.
Equation~\eqref{eq:trustee-threshold-derivation} compares immediate rewards;
the learned trustee policy also depends on continuation values, the
type-specific discount $\gamma_\phi$ and exploration.

\subsection{Linear response of the anchored loop}
\label{app:feedback-derivation}
The drive vanishes at the reference observables, while finite-time baseline
drift can supply a nonzero input after the checkpoint.
With binary unit investments, the population-mean investor payoff in any round is exactly
\begin{equation}
 \bar r_I(t)=\left(\frac{m(t)}2-1\right)f_{11}(t)-f_{10}(t).
 \label{eq:mean-investor-payoff-decomposition}
\end{equation}
Thus the cooperation-rate term satisfies
$\partial E(t+1)/\partial C(t)=\nu\beta_C$ when the payoff input is held fixed.
If $f_{11}$ changes at fixed $m$ and $f_{10}$, the payoff channel contributes
\begin{equation}
 \left.\frac{\partial\bar r_I}{\partial f_{11}}\right|_{m,f_{10}}
 =\frac m2-1,
 \end{equation}
so the combined one-round drive slope along that direction is
$\nu[\beta_C+\beta_r(m/2-1)]$. Other reallocations of outcome mass give
different directional derivatives; for example, converting $10$ to $11$ at a
fixed investment rate changes the payoff by $m/2$ per unit mass. The perturbation direction therefore determines how the two drive terms combine.

For a fixed analysis window, define the ensemble static-baseline responses
\begin{equation}
 F_{\mathcal W}(m,\rho)=\mathbb E_s[\bar f_{11,s}],\qquad
 R_{\mathcal W}(m,\rho)=\mathbb E_s[\bar r_{I,s}],
\end{equation}
with the population size, initialization and analysis window held fixed.
The local approximation uses smooth proxies for the responses generated by
the hard observation bins and $\varepsilon$-greedy action rule.
Let $\widetilde F_{\mathcal W}$ and
$\widetilde H_{\mathcal W}$ be smooth proxies for the mean $f_{11}$ and
$f_{10}$ responses within unchanged payoff and observation tiers.
Define the payoff proxy consistently with Eq.~\eqref{eq:mean-investor-payoff-decomposition}:
\[
 \widetilde R_{\mathcal W}
 =\left(\frac m2-1\right)\widetilde F_{\mathcal W}
 -\widetilde H_{\mathcal W}.
\]
The formal susceptibilities are
$\chi_{C,m}^{\mathcal W}=\partial\widetilde F_{\mathcal W}/\partial m$
and $\chi_{r,m}^{\mathcal W}=\partial\widetilde R_{\mathcal W}/\partial m$,
evaluated at $(m_0,\rho_0)$ within such a region.
Equation~\eqref{eq:env-to-m} gives
\begin{equation}
 M'(E_0):=\left.\frac{dm}{dE}\right|_{E_0}
 =k(m_{\max}-m_{\min})z_0(1-z_0),\qquad
 z_0=\frac{m_0-m_{\min}}{m_{\max}-m_{\min}}.
 \label{eq:environment-map-slope}
\end{equation}
Within the fixed payoff and observation tiers, the quasi-static scalar linearization reads
\begin{align}
 \delta E(t+1)&\simeq
 \left[1-\nu+g_{\mathcal W}(m_0,\rho_0)\right]\delta E(t),
 \label{eq:environment-perturbation}\\
 g_{\mathcal W}&=\nu M'(E_0)
 \left(\beta_C\chi_{C,m}^{\mathcal W}
 +\beta_r\chi_{r,m}^{\mathcal W}\right).
 \label{eq:finite-horizon-loop-gain}
\end{align}
Equation~\eqref{eq:environment-perturbation} governs a small \emph{perturbation}
$\delta E$ about the quasi-static operating point. Writing
$G_{\mathcal W}=g_{\mathcal W}/\nu$, local stability of this scalar map
requires $|1-\nu+\nu G_{\mathcal W}|<1$, or
$1-2/\nu<G_{\mathcal W}<1$. With a positive multiplier in this range,
the perturbation decays monotonically while collective outcomes can drive
changes in the operating level $E(t)$.

The payoff susceptibility can be decomposed directly from
Eq.~\eqref{eq:mean-investor-payoff-decomposition}:
\begin{equation}
 \chi_{r,m}^{\mathcal W}=\frac12\widetilde F_{\mathcal W}
 +\left(\frac{m_0}{2}-1\right)\chi_{C,m}^{\mathcal W}
 -\frac{\partial\widetilde H_{\mathcal W}}{\partial m}.
 \label{eq:payoff-susceptibility}
\end{equation}
This expression separates the mechanical return effect from behavioral shifts
between successful trust (11), untrustworthy outcomes (10), and
non-investment.

The simulated $\rho(E)$ map has zero derivative inside each tier and no
derivative at a boundary.
A crossing can instead be described over a finite interval by
$h'_{\rho,\mathrm{eff}}=\Delta\rho/\Delta E$.
At fixed $m$, define response secants over the same endpoints as
$\widehat\chi_{C,\rho}^{\mathcal W}=\Delta F_{\mathcal W}/\Delta\rho$
and $\widehat\chi_{r,\rho}^{\mathcal W}=\Delta R_{\mathcal W}/\Delta\rho$.
Their finite-crossing drive coefficient is
\begin{equation}
 g_{\rho,\mathcal W}=\nu h'_{\rho,\mathrm{eff}}
 \left(\beta_C\widehat\chi_{C,\rho}^{\mathcal W}
 +\beta_r\widehat\chi_{r,\rho}^{\mathcal W}\right).
\end{equation}
This coefficient describes a finite crossing between tiers;
Eqs.~\eqref{eq:environment-perturbation}--\eqref{eq:finite-horizon-loop-gain}
describe local response within a tier.

Within the smooth approximation, a steeper $\widetilde F_{\mathcal W}$
contributes a larger local gain.
The simulated learners also retain the preceding payoff history through
their $Q$-values and visited states.
Where a history-dependent linear-response approximation applies, response
kernels would therefore replace the static susceptibilities.
The scalar gain summarizes a quasi-static response within the
fast--slow feedback framework~\cite{Weitz2016,Tilman2020EnvFeedback}.

We construct the finite-difference benchmark for the $0.42$ comparison in
Section~\ref{sec:LoopGain} from baseline means on the 31-point grid at
$\rho_0=0.2075$, over $m\in[2.9132,5.7184]$.
Both this grid and the feedback slice use 20 seeds and window $[3000,12000)$.
We define $\widehat F_{\mathcal W}$ by piecewise-linear interpolation of the
grid means, with constant extension beyond either endpoint, and compute the slope as
\[
 \widehat\chi_{C,m}^{\mathcal W}(m_0;h)
 =\frac{\widehat F_{\mathcal W}(m_0+h,\rho_0)
       -\widehat F_{\mathcal W}(m_0-h,\rho_0)}{2h},
 \qquad h=0.25.
\]
The comparison retains nominal slice points with $m_0\ge3.6$ and distance
at least $0.1$ from either observation-tier edge.
This selection leaves 11 points based on their nominal centers.
Intervals centered at $m_0=4.5$ and $4.9$ cross the upper observation edge,
while those at $5.5$, $5.7$ and $5.9$ use endpoint extension.
At each selected point, the benchmark is
$\widehat\chi_{C,m}^{\mathcal W}\Delta\bar m$.
The 11 signed observed-to-benchmark ratios have median
$0.42$, interquartile range $[0.252,0.642]$
(linearly interpolated quartiles), and range $[-0.009,33.125]$.
At $m_0=5.9$, endpoint extension gives a small benchmark denominator
($-0.000117$) and a ratio of $33.125$; at $5.1$, a near-zero observed
response reverses sign relative to the benchmark. We include all 11 points
in these ratio summaries.

If an environmental perturbation follows this scalar map with a positive
multiplier, we calculate $G_{\mathrm{eff}}$ from the time constant $\tau$
of an exponential fit:
\begin{equation}
 G_{\mathrm{eff}}=\frac{e^{-1/\tau}-1+\nu}{\nu},
 \qquad
 G_{\mathrm{eff}}\simeq1-\frac{1}{\nu\tau}.
 \label{eq:gain-from-relaxation}
\end{equation}
The first expression is the exact discrete-time conversion; the second is
its continuous approximation. We infer $G_{\mathrm{eff}}$ from the fitted
environmental relaxation time, which includes contributions from mechanical
payoff feedback. The static-susceptibility expression in
Eq.~\eqref{eq:finite-horizon-loop-gain} defines the separate quasi-static benchmark.

\subsection{Onset estimate}
\label{app:onset}
For each parameter setting, $q_0$ is measured at the specified point
$m=3.0$ by pooling the raw outcome counts over the fixed analysis window,
\begin{equation}
 q_0=\frac{N_{11}}{N_{10}+N_{11}}.
 \label{eq:q0-count-estimator}
\end{equation}
This point is evaluated independently of the nine-point logistic fit and lies
well below its fitted midpoint. Substituting $q_0$ for $\hat q$ in
Eq.~\eqref{eq:investor-threshold-derivation} and solving
$q_0=2\lambda/(m+2\lambda-2)$ for $m$ gives
Eq.~\eqref{eq:m1-theory}. For uncertainty propagation, each bootstrap draw
resamples the 20 seed indices as whole units, refits the complete
$\bar f_{11}(m)$ curve, recomputes $q_0$ from the corresponding resampled
counts, and then evaluates the pooled midpoint's offset from the RS reference.
The nine combinations of discount factor and exploration rate in
Fig.~\ref{fig:theory-validation}(a) and the five loss-aversion settings in
Fig.~\ref{fig:theory-validation}(b) comprise 14 parameter settings, each of which
yielded 1,000 valid bootstrap draws. Logistic fits had $R^2$ values from
0.9366 to 0.9988.
The positive offset compares the pooled finite-horizon midpoint with the
RS one-round utility zero.

\section{State space and learning details}
\label{app:state-space}
All continuous state signals are discretized with the same three-level map
\begin{equation}
 b_3(x)=
 \begin{cases}
 0, & 0\leq x<1/3,\\
 1, & 1/3\leq x<2/3,\\
 2, & 2/3\leq x\leq1,
 \end{cases}
 \label{eq:bin3}
\end{equation}
after clipping $x$ to $[0,1]$. Labels $0,1,2$ correspond to low, medium, and
high. The public environment component bins the branch-specific latent signal
$m_{\rm signal}(E_t)$ through
\begin{equation}
 z_{\rm signal}(t)=\operatorname{clip}\!\left(
 \frac{m_{\rm signal}(E_t)-m_{\min}}{m_{\max}-m_{\min}},0,1\right),\qquad
 b_E(t)=b_3\!\left(z_{\rm signal}(t)\right).
 \label{eq:environment-bin}
\end{equation}
All feedback branches use the same signal mapping,
$z_{\rm signal}(t)=\sigma(kE_t)$ by Eq.~\eqref{eq:env-to-m}, evaluated along
their own environmental trajectories. In the $\rho$-only branch, the realized
  payoff multiplication factor remains $m_0$, while the public tier is computed from
that branch's latent signal. In the static baseline,
$m_{\rm signal}=m_0$ and the corresponding tier is constant.

After investor $i$ is matched to trustee $j(i,t)$, its decision state is
\begin{equation}
 s_i(t)=\bigl(X_i(t-1),\ b_3(T_{i,j(i,t)}(t)),\ b_E(t)\bigr).
 \label{eq:investor-state}
\end{equation}
Here $X_i(t-1)\in\{00,01,10,11\}$ is the previous public outcome code and
$T_{i,j(i,t)}$ is the pair-specific trust weight for the current match.
Trustee actions are public in the model even when the matched investor
does not invest, so both $00$ and $01$ are observable outcome codes. There are
$4\times3\times3=36$ investor states and two actions,
$A_I=\{0,1\}$, giving a $36\times2$ table for each investor. Investor type
enters through the subjective reward in Eq.~\eqref{eq:utility-investor}.
Both actions are available in every state for all investor types.

Trustees observe aggregated inflow before acting. With
$\mathcal M_j(t)=\{i:j(i,t)=j\}$ and $n_j(t)=|\mathcal M_j(t)|$, define
\begin{equation}
 I_j(t)=\sum_{i\in\mathcal M_j(t)}a_i(t),\qquad
 \widetilde I_j(t)=
 \begin{cases}
 I_j(t)/n_j(t), & n_j(t)>0,\\
 0, & n_j(t)=0,
 \end{cases}
 \label{eq:normalized-inflow}
\end{equation}
where the stake is one unit, so $\widetilde I_j$ is the invested share
of trustee $j$'s current matches. Its decision state is
\begin{equation}
 \begin{aligned}
 s_j(t)=\bigl(&b_3(R_j(t)),\ b_3(\widetilde I_j(t)),\ a_j(t-1),\\
              &b_3(\overline T_j(t)),\ b_E(t)\bigr),\\
 \overline T_j(t)&=\frac1{N_I}\sum_{i=1}^{N_I}T_{ij}(t).
 \end{aligned}
 \label{eq:trustee-state}
\end{equation}
The component cardinalities are $3\times3\times2\times3\times3=162$.
Together with $A_T=\{U,T\}$, each trustee therefore has a
$162\times2$ table.

The continuous memory variables remain in $[0,1]$ under the updates
\begin{align}
 R_j(t+1)&=(1-\eta_R)R_j(t)+\eta_R
 \mathbf1[I_j(t)>0]\mathbf1[a_j(t)=T],
 \label{eq:rep-update-app}\\
 T_{ij}(t+1)&=(1-\eta_T)T_{ij}(t)+\eta_T
 \mathbf1[a_j(t)=T]
 \quad\text{if }i\text{ invested with }j,
 \label{eq:trust-update-app}
\end{align}
and $T_{ij}(t+1)=T_{ij}(t)$ otherwise. Pairwise trust therefore changes
through realized investment interactions.

Algorithm~\ref{alg:environment-feedback} gives the round-by-round update order.
Both roles use the same environment tier $b_E(t)$ and complete their previous
transition once the current decision state is available.
The shared round-3000 checkpoint retains both $Q$-tables, all memories and
counters, previous states, actions and rewards, the environmental state,
seed-specific references, and random-number streams.
Each branch therefore continues from the same learned state, including the
transition awaiting its next $Q$ update.

\clearpage
\section{Parameters and run specifications}
\label{app:parameters}
\suppressfloats[t]
\begin{table}[htbp]
\centering
\caption{Simulation settings for Figs.~\ref{fig:model-schematic}--\ref{fig:feedback-sensitivity-grid} and Supplementary
Figs.~S1--S3. Feedback runs use the centered law, a seed-specific historical
reference, and a shared round-3000 checkpoint. Figure intervals use the seed units listed for each run.}
\label{tab:run-specifications}
\scriptsize
\setlength{\tabcolsep}{1.5pt}
\renewcommand{\arraystretch}{0.96}
\begin{tabular}{@{}L{0.09\textwidth}L{0.205\textwidth}L{0.125\textwidth}L{0.16\textwidth}L{0.07\textwidth}L{0.245\textwidth}@{}}
\toprule
Figure & Experiment & $(N_I,N_T)$ & Horizon; window & Seeds & Conditions \\
\midrule
Fig.~\ref{fig:model-schematic} & centered-$E_0$ schematic & --- & --- & --- & --- \\
Fig.~\ref{fig:main-result-overview}\allowbreak(a,c) & fixed-$\rho_0$ feedback comparisons & $(240,120)$ & 12k; $[3\mathrm{k},12\mathrm{k})$ & 0--19 & same data as Figs.~\ref{fig:feedback-landscape}c and \ref{fig:channel-control} \\
Fig.~\ref{fig:main-result-overview}\allowbreak(b) & trust trajectories at D & $(240,120)$ & 12k; $[3\mathrm{k},12\mathrm{k})$ & 100--149 & same data as Fig.~\ref{fig:representative-points}l \\
Fig.~\ref{fig:baseline-phase-panel} & baseline grids & $(240,120)$ & 12k; $[3\mathrm{k},12\mathrm{k})$ & 0--19 & static baseline \\
Fig.~\ref{fig:timeseries-parameter-grid} & six-$m$ time series & $(240,120)$ & 12k; $[3\mathrm{k},12\mathrm{k})$ & 0--19 & static baseline \\
Fig.~\ref{fig:feedback-landscape}\allowbreak(a,b,d) & feedback grid & $(240,120)$ & 12k; $[3\mathrm{k},12\mathrm{k})$ & 0--19 & baseline, $m$-only \\
Fig.~\ref{fig:feedback-landscape}\allowbreak(c) & fixed-$\rho_0$ slice & $(240,120)$ & 12k; $[3\mathrm{k},12\mathrm{k})$ & 0--19 & baseline, $m$-only, $\rho$-only, $m{+}\rho$ \\
Fig.~\ref{fig:representative-points} & A--D comparisons & $(240,120)$ & 12k; $[3\mathrm{k},12\mathrm{k})$ & 100--149 & baseline, $m$-only \\
Fig.~\ref{fig:mechanism-atlas} & mechanism atlas & $(240,120)$ & 12k; $[3\mathrm{k},12\mathrm{k})$ & 0--19 & static baseline \\
Fig.~\ref{fig:theory-validation} & parameter dependence of the fitted onset & $(240,120)$ & 12k; $[3\mathrm{k},12\mathrm{k})$ & 0--19 & baseline; bootstrap $B=1000$ \\
Fig.~\ref{fig:selected4-neti-matrix-grid} & A--D NETI matrices & $(240,120)$ & 12k; $[3\mathrm{k},12\mathrm{k})$ & 100--149 & baseline, $m$-only; paired \\
Fig.~\ref{fig:feedback-controls}\allowbreak(a) & schedule-yoked & $(240,120)$ & 12k; $[3\mathrm{k},12\mathrm{k})$ & 200--219 & baseline, $m$-only, yoked \\
Fig.~\ref{fig:feedback-controls}\allowbreak(b) & channel comparisons at A and D & $(240,120)$ & 12k; $[3\mathrm{k},12\mathrm{k})$ & 100--149 & baseline, $m$-only, $\rho$-only, $m{+}\rho$ \\
Fig.~\ref{fig:feedback-controls}\allowbreak(c,e,f) & horizon and windows at A and D & $(240,120)$ & c: 12k, 24k, 36k; $[3\mathrm{k},T)$; e: 12k, refs $[1\mathrm{k},3\mathrm{k})$ or $[2\mathrm{k},3\mathrm{k})$; f: 12k, obs. $[3\mathrm{k},12\mathrm{k})$ or $[6\mathrm{k},12\mathrm{k})$ & 200--219 & baseline, $m$-only \\
Fig.~\ref{fig:feedback-controls}\allowbreak(d) & baseline finite size & $(60,30)$ to $(480,240)$ & 12k; $[3\mathrm{k},12\mathrm{k})$ & 0--19 & baseline; bootstrap $B=1000$ \\
Fig.~\ref{fig:feedback-sensitivity-grid} & sensitivity grid at A and D & $(240,120)$ & 12k; $[3\mathrm{k},12\mathrm{k})$ & 200--219 & baseline, $m$-only; $4\!\times\!4$ drive, $\nu$, noise, reference window \\
Fig. S1 & baseline horizon diagnostic & $(240,120)$ & 36k; three 3k blocks & 0--9 & static baseline \\
Fig. S2 & initial-condition control & $(240,120)$ & 12k; $[3\mathrm{k},12\mathrm{k})$ & 0--19 & paired low and high initial states \\
Fig. S3 & fixed-$\rho_0$ channels & $(240,120)$ & 12k; $[3\mathrm{k},12\mathrm{k})$ & 0--19 & baseline, $m$-only, $\rho$-only, $m{+}\rho$ \\
\bottomrule
\end{tabular}
\end{table}

We calculated within-seed feedback effects from shared checkpoints and random
streams (Table~\ref{tab:run-specifications}). We also compared the A--D effects
between two independent numerical implementations using separate seed ensembles.
Pointwise Welch intervals for differences in their $m$-only-minus-baseline
effects all included zero and spanned differences comparable to the feedback
effects, especially at C. The source data provide the four estimates and intervals.

For schedule-yoked replay, we index the 20 recipient seeds as
$i=0,\ldots,19$; recipient $i$ receives
the schedule from donor $(i+1)\bmod20$. A contrast involving replay thus
depends on both seeds. We split the cycle into fixed even- and odd-recipient
groups, each containing ten disjoint recipient--donor pairs. Assuming
independent seed streams, pairs within either group are independent,
although the two groups share seeds. For group $g$, we form a two-sided
97.5\% approximate $t_9$ interval $[L_g,U_g]$ for its mean contrast. We report
\[
 \widehat\mu=\frac{\bar d_{\rm even}+\bar d_{\rm odd}}2,
 \qquad
 \left[\frac{L_{\rm even}+L_{\rm odd}}2,
       \frac{U_{\rm even}+U_{\rm odd}}2\right].
\]
Bonferroni's inequality gives at least 95\% joint coverage when the marginal
intervals attain their nominal coverage; averaging endpoints then covers
the overall mean. Coverage remains approximate because each ten-pair
interval uses a Student-$t$ approximation. This pointwise procedure uses
all 20 contrasts and accounts for shared donors without assuming
independence between groups. The $m$-only-minus-baseline contrast involves
no donor and uses the ordinary paired $t_{19}$ interval.

\section{Supplementary numerical experiments}
\label{app:supp-experiments}
\setcounter{figure}{0}
\renewcommand{\thefigure}{S\arabic{figure}}
\subsection{Trust dynamics over longer horizons}

We extended the baseline runs to 36,000 rounds at three specified points,
using 10 seeds. Each 12k, 24k, and 36k checkpoint summarizes the preceding
3,000 rounds within each seed. The convergence gate required an absolute mean
36k-minus-24k change no larger than 0.01 and a paired 95\% Student-$t$
interval containing zero.

\begin{figure}[!htbp]
  \centering
     \includegraphics[width=1\textwidth]{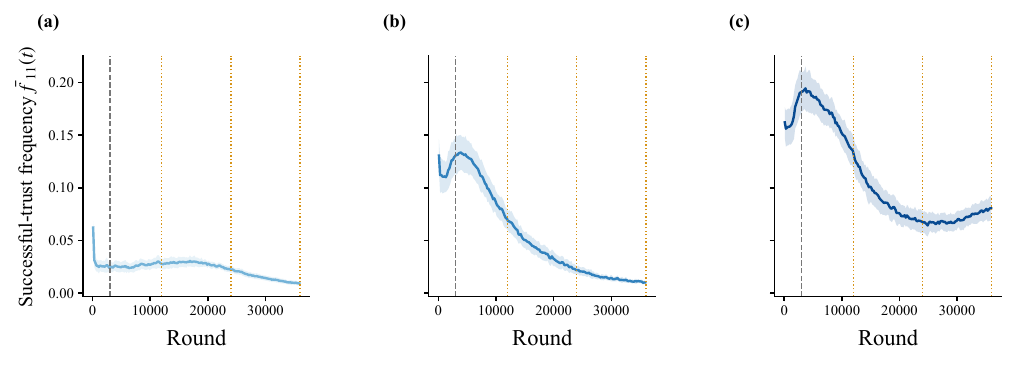}
  \caption{Baseline trust dynamics over 36,000 rounds at $\rho=0.2075$.
  Panels (a--c) use $m=3.2$, 4.6, and 5.6, respectively.
  Curves show 200-round block means and two-sided 95\% Student-$t$
  confidence intervals across 10 static-baseline seeds; the dashed line marks
  the burn-in cutoff and the dotted lines the checkpoints at 12,000, 24,000,
  and 36,000 rounds.
  Successful trust continued to decline at $m=3.2$ and 4.6, while $m=5.6$
  showed a late reversal.}
  \label{fig:convergence-diagnostic}
\end{figure}

\begin{table}[!htbp]
\centering
\caption{Checkpoint means and convergence-gate decisions. The uncertainty
on $\Delta_{36-24}$ is the paired 95\% confidence-interval half-width.}
\label{tab:convergence-gate}
\begin{tabular}{@{}rrrrrc@{}}
\toprule
$m$ & 12k & 24k & 36k & $\Delta_{36-24}$ & Gate \\
\midrule
3.2 & 0.02802 & 0.02462 & 0.01017 & $-0.01445\pm0.00280$ & fail \\
4.6 & 0.08233 & 0.02588 & 0.01103 & $-0.01485\pm0.00380$ & fail \\
5.6 & 0.14789 & 0.07025 & 0.07794 & $+0.00768\pm0.00628$ & fail \\
\bottomrule
\end{tabular}
\end{table}

All three points failed the gate because their paired intervals excluded zero.
The first two continued to decline, whereas the high-$m$ point reversed upward
late in the run. The 12,000-round estimates therefore describe behavior
during ongoing learning.

\subsection{Representative-point and NETI uncertainty}

Table~\ref{tab:selected-neti-ci} summarizes the paired feedback effects and
two NETI-cell contrasts from Figs.~\ref{fig:representative-points} and
\ref{fig:selected4-neti-matrix-grid}. Each NETI cell measures finite-window
joint-frequency excess relative to its independent-marginal benchmark.

\begin{table}[!htbp]
\centering
\caption{Feedback effects and paired NETI differences at A--D. Values are
means $\pm$ 95\% Student-$t$ confidence-interval half-widths over 50 matched
seeds and the fixed analysis window.}
\label{tab:selected-neti-ci}
\small
\begin{tabular}{@{}lrrr@{}}
\toprule
Point & $\Delta\bar f_{11}$ & $\Delta$NETI $10\!\to\!11$ & $\Delta$NETI $11\!\to\!11$ \\
\midrule
A & $+0.000271\pm0.000430$ & $+0.000016\pm0.000033$ & $+0.000007\pm0.000033$ \\
B & $-0.004545\pm0.002173$ & $-0.000694\pm0.000462$ & $-0.000553\pm0.000363$ \\
C & $-0.003114\pm0.005335$ & $-0.001583\pm0.000659$ & $-0.001148\pm0.000580$ \\
D & $-0.004022\pm0.001615$ & $-0.000347\pm0.000230$ & $-0.000190\pm0.000230$ \\
\bottomrule
\end{tabular}
\end{table}

\subsection{Finite-horizon initial-condition dependence}

\begin{figure}[!htbp]
  \centering
       \includegraphics[width=1\textwidth]{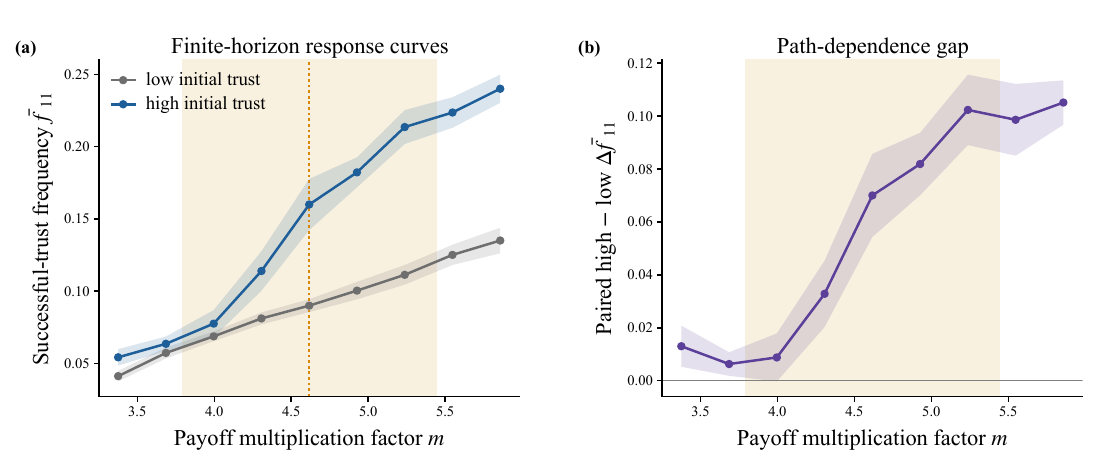}
  \caption{Finite-horizon initial-condition control at $\rho=0.2075$.
  (a) Successful-trust frequency for paired low- and high-trust
  initializations. The shaded region marks the fitted crossover band, and the
  dotted line marks its midpoint (fitted $m^*=4.62$, $w=0.83$ at this
  $\rho$). (b) Paired high-minus-low difference. Curves and bands show means
  and paired 95\% Student-$t$ confidence intervals across 20 matched
  response seeds of the static baseline over the fixed $[3000,12000)$
  analysis window.}
  \label{fig:initial-condition-control}
\end{figure}

We paired low- and high-trust initializations across 20 response seeds at
$\rho=0.2075$ and nine specified payoff multiplication factors around the fitted
baseline crossover. We set $T_{ij}(0)=R_j(0)=0.1$ in the low-trust
initialization and $0.9$ in the high-trust initialization.
We set $X_i(-1)=00$ and $a_j(-1)=U$ in the low-trust initialization,
and $X_i(-1)=11$ and $a_j(-1)=T$ in the high-trust initialization.
Both initializations used zero $Q$-values and the same seed within each pair.
Figure~\ref{fig:initial-condition-control} shows the
response curves and their paired gap. The high-minus-low gap was positive
with a 95\% Student-$t$ interval excluding zero at eight of nine sampled
points. The interval at $m=3.9963$ included zero.
At the fitted midpoint $m=4.6168$, the gap was
$+0.07000\pm0.01569$; it reached $+0.10509\pm0.00838$ at the largest
sampled $m$.

These paired trajectories at fixed parameters establish initial-condition
dependence over the specified finite window.

\subsection{Centered payoff-channel control}

\begin{figure}[!htbp]
  \centering

         \includegraphics[width=1\textwidth]{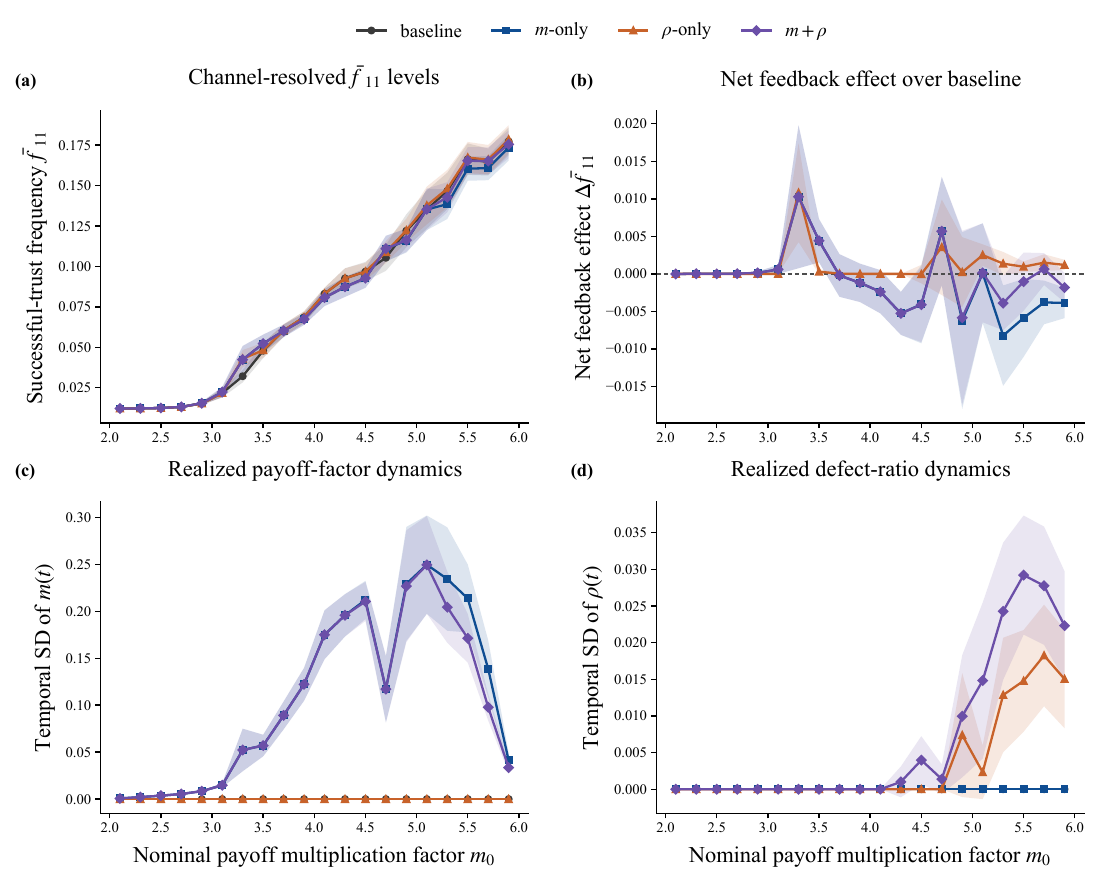}
  \caption{Centered payoff-channel controls at fixed
  $\rho_0=0.2075$. Colors and markers identify the same branches throughout.
  (a) Successful-trust levels for baseline, $m$-only,
  $\rho$-only, and joint-channel branches. (b) Within-seed branch-minus-
  baseline changes; the $\rho$-only branch produces a positive response at
  $m_0=3.3$ while retaining both nominal payoff parameters. The horizontal
  dashed line in (b) marks zero effect. (c) Temporal standard deviation of
  realized $m(t)$; it is zero in the baseline and $\rho$-only branches,
  where $m(t)$ remains fixed. (d) Temporal standard deviation of realized
  $\rho(t)$; it is zero in the baseline and $m$-only branches,
  where $\rho(t)$ remains fixed. Lines and bands report means and pointwise 95\%
  Student-$t$ confidence intervals across 20 matched seeds.}
  \label{fig:channel-control}
\end{figure}

We compared the baseline, $m$-only, $\rho$-only, and joint branches
across 20 nominal $m$ values at $\rho_0=0.2075$, using the same 20 seeds
(Fig.~\ref{fig:channel-control}). Table~\ref{tab:slice-channels}
lists a selection of nine points from this slice. Across the slice, mean $m$-only changes in
$f_{11}$ ranged from $-0.0082$ to $+0.0103$. The corresponding ranges were
$-0.00004$ to $+0.0108$ for $\rho$-only and $-0.0058$ to $+0.0103$ for
the joint branch.

Nonzero $\rho$-only point estimates occurred at sampled $m_0=3.3$--$3.7$
and $4.5$--$5.9$, although some intervals included zero. Through the sampled
$m_0=4.7$ point, this branch retained both nominal payoff parameters and
served as a perception control. From $m_0=4.9$ upward, realized $\rho$
could also vary, adding payoff changes to the perception response.

The payoff factor $m(t)$ remained constant in the baseline and $\rho$-only
branches, while $\rho(t)$ remained constant in the baseline and $m$-only branches.
Each feedback branch followed its own environmental and perceived-state trajectory.

\section{Channel and discretization controls}
\label{app:identification}

We tested environmental relaxation, channel effects, continuation value and
state discretization. The impulse probe
(\ref{app:ident-amplifier}) and $2\times2$ payoff$\times$perception factorial
(\ref{app:ident-channels}) compare paired groups initialized from the same learned state at round 3000
using common random numbers. The discount-factor
(\ref{app:ident-myopic}) and discretization (\ref{app:ident-edges}) controls
change parameters used during learning. For these comparisons, we used matched seeds and separate burn-in periods
for each parameter setting, yielding different learned states at round 3000.

Unless stated otherwise, outcome means use
$\mathcal W_{\rm full}=[3000,12000)$. The channel time series also reports
$\mathcal W_{\rm late}=[10000,12000)$, and the impulse fits use the specified
post-shock interval. Reused seed identifiers form blocks across operating
points and arms. We report each point estimate followed by one standard error.
For simple contrasts, this is the sample standard deviation across independent
seed blocks divided by $\sqrt n$. For fitted or selection-dependent summaries, standard errors come from a
delete-one-seed jackknife. We construct pointwise $95\%$ confidence intervals
using a Student-$t$ approximation with $n-1$ degrees of freedom and regard a
signed effect as resolved when its interval excludes zero. Contrast definitions and summary
results appear below and in Table~\ref{tab:identification-summary}.

\subsection{Environmental relaxation after a small impulse}
\label{app:ident-amplifier}

Starting from the same learned state and using common random numbers,
we imposed positive and negative environmental displacements $\pm\delta E$ ($\delta E=0.05$) at round $T_0=7000$.
At the mid-band operating point, no tier edge was crossed, so the dynamic
perception bins remained at their nominal value. Both payoff parameters varied with the environment, while the perceived
environmental tier remained unchanged. For each of ten
seeds (100--109), we formed the symmetrized response
$h_s(k)=[m_{+,s}(T_0+k)-m_{-,s}(T_0+k)]/2$ and subtracted its mean over
$k\in[1800,2500)$.

We fitted the seed-mean curve to a single exponential using weighted
log-linear regression with squared-residual weights $h(k)^2$. The fit used
$k\in[5,200)$ and stopped when the response first fell below
$\max(0.03h_{\rm peak},0.0015)$. Here $h_{\rm peak}$ is the mean of the
first three response values. The fit required at least 12 points and used
$k=5,\ldots,121$ at the mid-band point. The reported $r^2$ uses the same
squared-residual weights in log space. We obtained the standard error of
$\tau$ by omitting each seed and repeating the fit.

\begin{figure}[!htbp]
  \centering
           \includegraphics[width=1\textwidth]{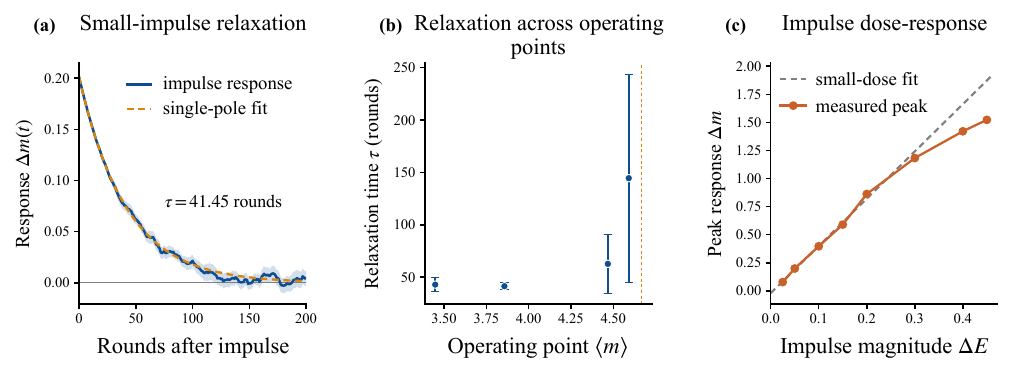}
  \caption{Environmental relaxation and response to different impulse amplitudes.
  We compared positive and negative perturbations from the same learned state
  using common random numbers. The environmental tier remained nominal at
  the mid-band operating point.
  (a) Symmetrized response of the operating point $m(t)$ to a small
  $\pm\delta E=0.05$ impulse at the mid-band operating point
  $\langle m\rangle\!\approx\!3.86$; the dashed curve is the single-pole fit
  $A\,e^{-t/\tau}$ ($\tau=41.45$ rounds, $r^2=0.997$). The line is the
  offset-corrected ten-seed mean and the band its
  $\pm1$ standard error. (b) Fitted relaxation time $\tau$ versus operating
  point (markers with $\pm1$ delete-one-seed jackknife standard error); the
  orange dotted line marks the boundary between the middle and high
  perception tiers at $m=14/3\simeq4.667$, near which the single-exponential
  fit deteriorated. (c) Peak displacement
  versus impulse magnitude with a small-signal linear fit (slope $4.21$,
  $r^2=0.998$); the response saturates gently at larger dose. All contrasts are
  within-seed paired.}
  \label{fig:ii3-amplifier}
\end{figure}

At the mid-band operating point ($\langle m\rangle\!\approx\!3.86$), the
fitted time constant was $\tau=41.45\pm3.51$ rounds
(approximate 95\% jackknife-$t_9$ CI $[33.51,49.38]$; $r^2=0.997$;
Fig.~\ref{fig:ii3-amplifier}a). Using the unrounded fitted $\tau$ and
$\nu=0.03$, Eq.~\eqref{eq:gain-from-relaxation} gives
$G_{\mathrm{eff}}=0.205$ in discrete time
(approximate 95\% CI $[0.020,0.332]$) and
$0.196$ in the continuous approximation
($[0.005,0.325]$). We obtain these intervals by monotonically
transforming the two endpoints of the $\tau$ interval under the scalar model.

The initial response divided by the imposed displacement was $3.92$, close
to the maximum analytic map slope $4.0$. Positive and negative impulse peaks
had a magnitude ratio of $1.01$. Peak displacement grew approximately
linearly with dose at small amplitude (slope $4.21$, $r^2=0.998$).
At $\delta E=0.45$, it reached $0.81$ of the linear extrapolation
(Fig.~\ref{fig:ii3-amplifier}c). This saturation is consistent with the
bounded map $m(E)$, whose range is $(2,6)$. As the operating point approached
the upper tier edge, the single-exponential fit deteriorated. The fitted
$\tau$ increased ($41.45\to62.8\to144.6$ rounds as
$\langle m\rangle$ moved $3.86\to4.47\to4.59$;
Fig.~\ref{fig:ii3-amplifier}b).

This fit characterizes environmental relaxation at the tested operating
point, including the mechanical payoff contribution described in
\ref{app:feedback-derivation}.

\subsection{Factorial payoff and perception controls}
\label{app:ident-channels}

We compared four combinations of dynamic or fixed payoff parameters and
perceived environmental tiers in a $2\times2$ factorial design.
Each comparison started from the same learned state and used common random
numbers, with payoff factor $P$ and perception factor $S$.
The payoff-on setting varied both $m$ and $\rho$; the main-text $m$-only
branch varied only the payoff multiplication factor. The four combinations $c$ were defined as follows:
\begin{center}
\small
\begin{tabular}{@{}lll@{}}
\toprule
Cell & Realized payoff parameters & Perceived environmental tier \\
\midrule
$P_+S_+$ & $m_{\rm signal}(E_t^c),\ \rho_{\rm dyn}(E_t^c)$ & dynamic, $b_E(E_t^c)$ \\
$P_+S_-$ & $m_{\rm signal}(E_t^c),\ \rho_{\rm dyn}(E_t^c)$ & fixed at its nominal value \\
$P_-S_+$ & $m_0,\ \rho_0$ & dynamic, $b_E(E_t^c)$ \\
$P_-S_-$ & $m_0,\ \rho_0$ & fixed at its nominal value \\
\bottomrule
\end{tabular}
\end{center}
Each cell updates its own $E_t^c$ from $C_t^c$ and $\bar r_{I,t}^c$
using Eq.~\eqref{eq:env-update}, with shared initial state and reference
values. In $P_-S_-$, the latent environment continues to update while both
routes to the agents remain closed. Common random numbers preserve the
pairing but allow realized environment paths to diverge. We compared successful trust across the four combinations to estimate
payoff and perception effects and their interaction, including the
subsequent environmental changes within each combination.

The operating points are $m_0=3.3$, $4.0$ and $4.7$, each with seeds
100--109. Let $Y_{ab,s,j}^{W}$ be the successful trust frequency in cell
$(a,b)$, seed $s$ and operating point $j$, averaged over window $W$; in the
following expressions the seed and point indices are suppressed:
\[
\begin{aligned}
 d_P^{W}&=\tfrac12[(Y_{++}^{W}-Y_{-+}^{W})+(Y_{+-}^{W}-Y_{--}^{W})],\\
 d_S^{W}&=\tfrac12[(Y_{++}^{W}-Y_{+-}^{W})+(Y_{-+}^{W}-Y_{--}^{W})],\\
 i^{W}&=Y_{++}^{W}-Y_{+-}^{W}-Y_{-+}^{W}+Y_{--}^{W}.
\end{aligned}
\]
For the two boundary points $j\in\{L,U\}$, we distinguish the signed average
and the mean absolute response:
\[
 \begin{aligned}
 u_s^W&=\tfrac12(d_{S,s,L}^W+d_{S,s,U}^W), &
 a_s^W&=\tfrac12(|d_{S,s,L}^W|+|d_{S,s,U}^W|),\\
 \widehat\mu_W&=\frac1{10}\sum_su_s^W, &
 \widehat A_W&=\frac1{10}\sum_sa_s^W.
 \end{aligned}
\]
For the interaction averaged over the three fixed operating points, we first
form $\bar i_s^W=\frac13\sum_ji_{s,j}^W$ and then average over seeds.
Standard errors and $t_9$ intervals use these ten seed-block values,
preserving dependence among the 20 or 30 seed--point cells.

\begin{figure}[!htbp]
  \centering
             \includegraphics[width=1\textwidth]{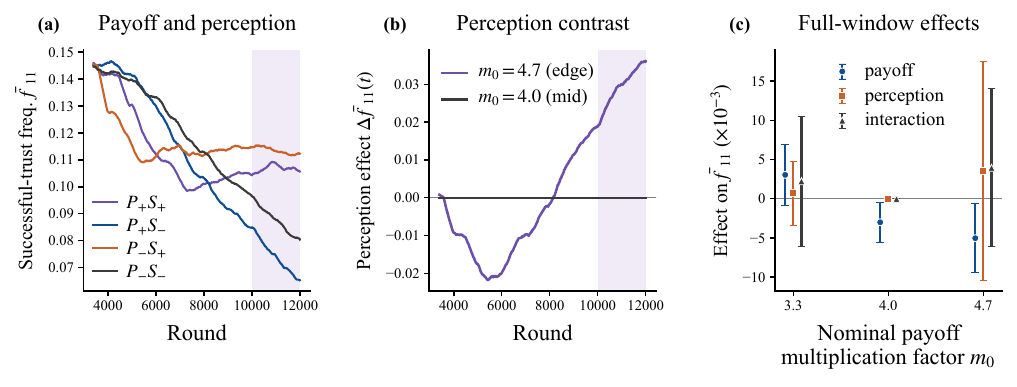}
  
  \caption{Payoff and perception effects in a $2\times2$ factorial comparison
  initialized from the same learned state using common random numbers and ten
  matched seeds per point. $P_+$ and $P_-$ denote conditions with environmental
  payoff feedback enabled and disabled, respectively; $S_+$ and $S_-$ denote
  dynamic and fixed perceived environmental tiers, respectively.
  The $P_+$ cells vary both $m$ and $\rho$, and each cell evolves its own
  environment. (a) Four-cell successful trust frequency
  $\bar f_{11}$ ($400$-round
  rolling mean of the seed average) at the upper edge $m_0=4.7$; shading in
  (a,b) marks $\mathcal W_{\rm late}=[10000,12000)$. (b) Perception contrast
  $\Delta\bar f_{11}(t)=\tfrac12[(P_{+}S_{+}-P_{+}S_{-})+(P_{-}S_{+}-P_{-}S_{-})]$:
  a build-up near the upper boundary ($m_0=4.7$) versus an exactly zero
  line in the mid-band ($m_0=4.0$). (c) Payoff and perception main effects and their interaction
  over $\mathcal W_{\rm full}=[3000,12000)$ versus $m_0$ (markers
  with pointwise $95\%$ $t_9$ intervals over ten seeds). At $m_0=4.0$, the
  observed perception main effect and interaction are both exactly zero.}
  \label{fig:ii2-channels}
\end{figure}

At the mid-band point, the payoff main effect was small but resolved,
$\Delta\bar f_{11}=-0.0030\pm0.0011$. Dynamic and frozen cells had identical successful-trust trajectories at both
payoff settings, giving zero perception and interaction contrasts
(Fig.~\ref{fig:ii2-channels}c).

At the two boundary points, the full-window mean absolute perception effect
was $\widehat A_{\rm full}=0.0098\pm0.0016$, whereas the
signed mean was $\widehat\mu_{\rm full}=+0.0021\pm0.0034$.
The per-point signed effects were $+0.0007\pm0.0018$ at the lower edge and
$+0.0036\pm0.0062$ at the upper, with both 95\% intervals including zero.
The time-resolved contrasts changed sign. At the upper boundary, the
initially negative contrast became positive in $\mathcal W_{\rm late}$,
$+0.0310\pm0.0063$ (Fig.~\ref{fig:ii2-channels}b). For this late window, the
corresponding two-boundary absolute mean was
$\widehat A_{\rm late}=0.0191\pm0.0025$.

The full-window payoff main effects were $+0.0030\pm0.0017$ at the lower
boundary, $-0.0030\pm0.0011$ mid-band and $-0.0050\pm0.0019$ at the upper
boundary. The lower-boundary $95\%$ interval $[-0.0008,+0.0069]$
included zero; the other two excluded zero. The interaction averaged across
all three points was $+0.0021\pm0.0021$ ($95\%$ CI
$[-0.0026,+0.0067]$, ten seed blocks). Its interval spanned zero and
included magnitudes comparable to the main effects.

\subsection{Continuation value and the RS fitted midpoint}
\label{app:ident-myopic}

Figure~\ref{fig:theory-validation} compares the pooled three-type midpoint
with the RS one-round reference in Eq.~\eqref{eq:m1-theory}.
We tested the role of continuation value by
comparing the learner with $\gamma_{\mathrm{inv}}=0.9$ against a
myopic control ($\gamma_{\mathrm{inv}}=0$), holding the other parameters
fixed. We fitted the risk-seeking type's $\bar f_{11}(m)$ response curves,
which yielded $r^2\approx0.98$--$0.99$.

For each discount factor, we fitted the risk-seeking response across the same $m$ grid.
We estimated uncertainty by deleting each of the six matched seeds from both
discount-factor settings and all grid points. Each deletion repeated the logistic fits and the
$q_0$ estimate at $\gamma_{\mathrm{inv}}=0.9$ at $m=3.0$. This jackknife propagated
covariance between the fitted midpoints and the common RS one-round
reference, $m_1^{\,*}=3.035\pm0.031$.

With the investor continuation value removed, the fitted RS midpoint was
$4.24\pm0.08$, a gap of $+1.21\pm0.08$ above the one-round
reference estimated at $\gamma_{\mathrm{inv}}=0.9$. Both groups continued to update
their $Q$-tables throughout the run. Restoring continuation value lowered the RS midpoint to
$3.61\pm0.04$, a paired shift of $-0.63\pm0.08$
($95\%$ approximate $t_5$ CI $[-0.85,-0.41]$).
This paired discount-factor contrast shows that continuation value advances
the RS crossover.

\begin{figure}[!htbp]
  \centering
             \includegraphics[width=1\textwidth]{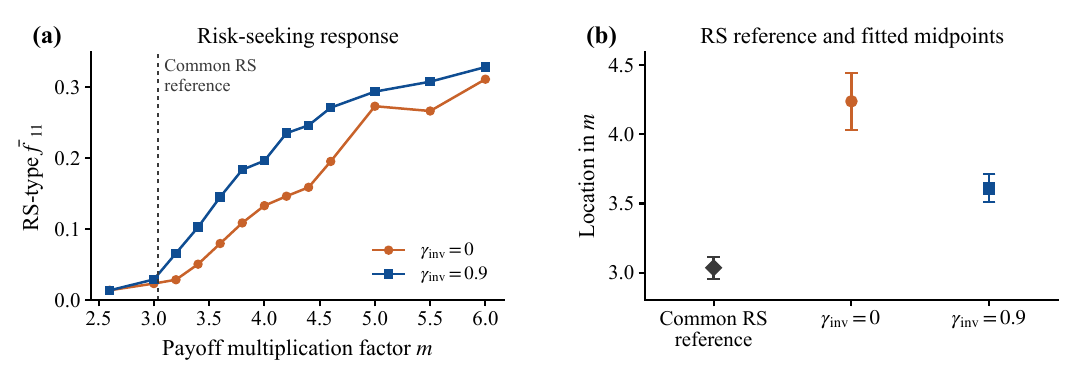}
  \caption{Dependence of the RS fitted midpoint on the discount factor.
  We varied only the investor discount factor $\gamma_{\mathrm{inv}}$ and
  used six matched seeds, with separate burn-in periods for each setting. (a) RS-type $\bar f_{11}(m)$ for the myopic
  ($\gamma_{\mathrm{inv}}=0$) and $\gamma_{\mathrm{inv}}=0.9$
  learners; the dashed line marks the common RS one-round reference
  ($m_1^{\,*}=3.03$).
  (b) The common reference, estimated at $\gamma_{\mathrm{inv}}=0.9$, and the two fitted midpoints. Error bars are approximate pointwise
  $95\%$ $t_5$ intervals from a matched-seed jackknife.}
  \label{fig:ii5-decomp}
\end{figure}

\subsection{Sensitivity to the common discretization thresholds}
\label{app:ident-edges}

We compared perception-switch effects under two sets of common
discretization thresholds.
The environmental edges satisfy $m_{\mathrm{edge}}=2+4\kappa$.
The schemes used the default thresholds $\kappa=(\tfrac13,\tfrac23)$
(edges at $3.333$ and $4.667$) and alternative thresholds $\kappa=(0.25,0.75)$
(edges at $3.000$ and $5.000$). For each threshold scheme, $m_0$, and seed, we started from the same learned
state and compared dynamic and fixed perceived environmental tiers.
Both $m$ and $\rho$ varied with the environment in each comparison.
The common $\kappa$ also defines the $\rho(E)$ tiers. Changing $\kappa$ moves
their switching boundaries from $E-E_0=\pm\ln2/k$ to $\pm\ln3/k$,
while retaining the three $\rho$ values.
For each matched pair, we measured the successful-trust contrast
$\mathrm{d}P\!E=\bar f_{11}(\text{dynamic})-\bar f_{11}(\text{frozen})$.

The paired cells shared $\kappa$, the payoff update rule, and the starting
values of all other $\kappa$-binned features (reputation, inflow, and trust).
The perception switch was the only difference imposed within each pair. While the dynamic
tier remains nominal, both cells present the same states to the learners.
A tier difference can change actions, subsequent payoffs and environmental
paths. Thus $\mathrm{d}P\!E$ measures the total closed-loop perception-switch
effect within a scheme. Across schemes, changing $\kappa$ jointly alters
state discretization, the $\rho$-switching rule and the independently generated
learning histories.

\begin{figure}[!htbp]
  \centering
               \includegraphics[width=1\textwidth]{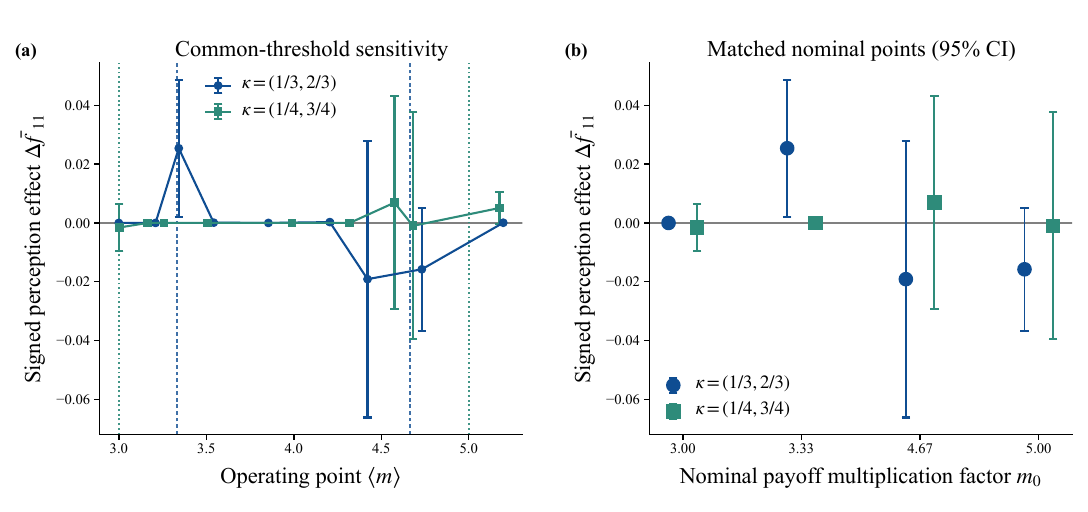}
  \caption{Perception responses under two sets of common discretization thresholds.
  We used six matched seeds per point and separate burn-in periods for each
  threshold scheme $\kappa$. (a) Signed mean perception contrast $\mathrm{d}P\!E$ versus
  operating point for the $\kappa=(\tfrac13,\tfrac23)$ scheme (edges at $3.33$ and $4.67$, dashed vertical lines)
  and the changed scheme (edges at $3.00$ and $5.00$, dotted vertical lines).
  (b) Comparisons at matched nominal $m_0$, including the nonzero-versus-zero
  contrast at $m_0=3.333$ and the less distinct upper-boundary results at
  $m_0=5.000$. Error bars in both panels are pointwise $95\%$ $t_5$ intervals.
  Both $m$- and $\rho$-feedback are active. The common $\kappa$ changes
  environmental and other state bins as well as the $\rho$-switching
  boundaries; the cross-scheme comparison measures their joint sensitivity.}
  \label{fig:ii6-edges}
\end{figure}

At $m_0=3.333$, near its lower boundary, the $\kappa=(\tfrac13,\tfrac23)$ scheme
gave $\mathrm{d}P\!E=+0.0254\pm0.0091$
($95\%$ $t_5$ CI $[+0.0021,+0.0488]$). The changed scheme produced
identical stored successful-trust trajectories in all six seed pairs.
At $m_0=3.000$, the changed scheme yielded a nonzero contrast and the
$\kappa=(\tfrac13,\tfrac23)$ scheme an observed zero. The upper-boundary comparison was less
distinct: at $m_0=5.000$, the changed scheme had a smaller absolute point
estimate and a wide interval. The $\kappa=(\tfrac13,\tfrac23)$ scheme gave a larger
absolute estimate (Fig.~\ref{fig:ii6-edges}).

For the proximity summary, each run's operating point was the frozen-perception
branch's full-window mean $m_{\rm signal}$. Absolute seed-level contrasts
pooled across schemes and nominal points averaged $0.0168\pm0.0032$
at distance at most $0.10$ from the nearest edge (17 seed--point contrasts).
They averaged $0.0056\pm0.0029$ at distance at least $0.30$
(56 contrasts). Standard errors deleted each of the six seeds across both
schemes and all points, preserving dependence from seed reuse.

The correlation of
absolute contrasts with proximity to the scheme's own edges was $+0.24$,
compared with $+0.01$ for the other scheme's edges. The change between nonzero
and zero perception contrasts near the lower edges demonstrates sensitivity
to the common discretization thresholds.

\begin{table}[!htbp]
\centering
\caption{Summary of the channel and discretization controls.
Uncertainties are one standard error across seed blocks or from a
delete-one-seed jackknife, as specified in the text. A dagger ($\dagger$)
marks signed effects whose pointwise $95\%$ Student-$t$ interval excludes
zero. ``Exact $0$'' denotes identical successful-trust trajectories.
Outcome windows are
$\mathcal W_{\rm full}=[3000,12000)$ and $\mathcal W_{\rm late}=[10000,12000)$.}
\label{tab:identification-summary}
\footnotesize
\setlength{\tabcolsep}{4pt}
\begin{tabularx}{\textwidth}{@{}L{0.21\textwidth}>{\raggedright\arraybackslash}XL{0.28\textwidth}@{}}
\toprule
Control & Quantity & Value \\
\midrule
Environmental impulse & mid-band relaxation time $\tau$ & $41.45\pm3.51$ rounds \\
(\ref{app:ident-amplifier}) & scalar $G_{\mathrm{eff}}$ in discrete time & $0.205$ \\
 & scalar $G_{\mathrm{eff}}$ in the continuous approximation & $0.196$ \\
 & initial response normalized by $\delta E$ (maximum map slope $4.0$) & $3.92$ \\
 & dose-response linear slope & $4.21$ ($r^2=0.998$) \\
\addlinespace
Channel $2\times2$ & full-window payoff effect, $m_0=4.0$ & $-0.0030\pm0.0011{}^{\dagger}$ \\
(\ref{app:ident-channels}) & perception main $+$ interaction, mid-band & exact $0$ \\
 & full-window boundary magnitude $\widehat A_{\rm full}$ & $0.0098\pm0.0016$ \\
 & full-window boundary signed mean $\widehat\mu_{\rm full}$ & $+0.0021\pm0.0034$ \\
 & late-window boundary magnitude $\widehat A_{\rm late}$ & $0.0191\pm0.0025$ \\
 & late-window signed perception effect, $m_0=4.7$ & $+0.0310\pm0.0063{}^{\dagger}$ \\
 & full-window interaction, three-point mean & $+0.0021\pm0.0021$ \\
\addlinespace
RS midpoint & common one-round reference $m_1^{\,*}$ & $3.035\pm0.031$ \\
(\ref{app:ident-myopic}) & myopic ($\gamma_{\mathrm{inv}}{=}0$) midpoint & $4.24\pm0.08$ \\
 & myopic midpoint minus common reference & $+1.21\pm0.08{}^{\dagger}$ \\
 & $\gamma_{\mathrm{inv}}{=}0.9$ midpoint & $3.61\pm0.04$ \\
 & continuation shift, $\gamma_{\mathrm{inv}}{:}0\!\to\!0.9$ & $-0.63\pm0.08{}^{\dagger}$ \\
\addlinespace
Discretization & $\mathrm{d}P\!E$ at $m_0{=}3.333$, $\kappa{=}(\tfrac13,\tfrac23)$ & $+0.0254\pm0.0091{}^{\dagger}$ \\
(\ref{app:ident-edges}) & same nominal point, changed scheme & exact $0$ \\
 & mean $|\mathrm{d}P\!E|$, distance $\le0.10$ & $0.0168\pm0.0032$ \\
 & mean $|\mathrm{d}P\!E|$, distance $\ge0.30$ & $0.0056\pm0.0029$ \\
 & corr$(|\mathrm{d}P\!E|,\text{proximity})$: own vs other & $+0.24$ vs $+0.01$ \\
\bottomrule
\end{tabularx}
\end{table}

\end{document}